\documentclass{aa}
\usepackage{graphicx}
\usepackage{natbib}
\usepackage{txfonts}

\def\hess{{HESS}}
\def\ls{LS~5039}
\def\lsi{LSI~+61$\degr$303}
\def\psrb{PSR~B1259-63}
\def\spose#1{\hbox to 0pt{#1\hss}}
\def\la{\mathrel{\spose{\lower 3pt\hbox{$\mathchar"218$}}
     \raise 2.0pt\hbox{$\mathchar"13C$}}}
\def\ga{\mathrel{\spose{\lower 3pt\hbox{$\mathchar"218$}}
     \raise 2.0pt\hbox{$\mathchar"13E$}}}

\begin{document}

\title{Gamma-ray binaries: pulsars in disguise ?}
\author{Guillaume Dubus
	\inst{1,2}
	}

\institute{Laboratoire Leprince-Ringuet, UMR 7638 CNRS, Ecole Polytechnique, 91128 Palaiseau, France; \email{dubus@in2p3.fr}
\and
Institut d'Astrophysique de Paris, UMR 7095 CNRS, Universit\'e Pierre \& Marie Curie, Paris 6, 98 bis bd Arago, 75014 Paris, France
}

\date{Draft \today}
\abstract
{\object{\ls}\ and \object{\lsi}\ are unique amongst high-mass X-ray binaries (HMXB) for their spatially-resolved radio emission and their counterpart at $>$GeV gamma-ray energies, canonically attributed to non-thermal particles in an accretion-powered relativistic jet. The only other HMXB known to emit very high energy (VHE) gamma-rays, \object{\psrb}, harbours a non-accreting millisecond pulsar.\\
The purpose is to investigate whether the interaction of the relativistic wind from a young pulsar with the wind from its stellar companion, as in \psrb, constitutes a viable scenario to explain the observations of \ls\ and \lsi. Emission arises from the shocked pulsar wind material, which then flows away to large distances in a comet-shape tail, reproducing on a smaller scale what is observed in isolated, high motion pulsars interacting with the interstellar medium.\\
The timescales for acceleration and radiation of particles at the shock between the pulsar wind and stellar wind are calculated. Simple expectations for the spectral energy distribution (SED) are derived and are shown to depend on few input parameters. Detailed modelling of the particle evolution is attempted and compared to the observations from radio to TeV energies.\\
Acceleration at the shock provides high energy electrons that steadily emit synchrotron in X-rays and inverse Compton scatter stellar light to $\gamma$-rays. Electrons streaming out of the system emit at IR frequencies and below. The {overall aspect of the} SEDs is {adequately} reproduced for standard values of the parameters. The morphology of the radio tail can mimic a microquasar jet. Good agreement is found with the published VLBI map of \ls\ and predictions are made on the expected change in appearance with orbital phase.\\
The pulsar wind scenario provides a common, viable framework to interpret the emission from all three $\gamma$-ray binaries.
}
\maketitle

\section{Introduction}
X-ray binaries are systems composed of a neutron star or black hole in orbit around a high or low-mass normal star (HMXBs and LMXBs). Non-thermal radio emission has been detected in X-ray binaries and is usually associated with outflows \citep[see][for a review]{fender_xrb}. In a few objects, {radio} observations are able to follow the propagation of radio emission up to parsec scales, after initial flaring in the central source. The inferred speeds of the collimated outflow are relativistic with bulk Lorentz factors $\Gamma$ of a few. These apparently superluminal ejection events have been tentatively associated with transitions from a hard to a soft X-ray spectrum \citep{fender}.

In many more objects, the radio emission has a flat spectrum and is correlated with a hard X-ray spectral state. Usually the radio emission is unresolved but {imaging by Very Long Baseline Interferometry (VLBI)} of Cyg X-1 in this state has revealed a compact, collimated radio outflow on a scale of a few AU \citep{stirling}. Asymmetries in the surface brightness probably reflect moderate relativistic boosting ($v/c\sim 0.3$). It is as yet unclear whether compact radio emission and superluminal ejections can be taken as different manifestations of a unique, underlying ejection mechanism. 

Similar types of radio emission (compact or superluminal ejections) are seen in active galactic nuclei (AGN), prompting speculations that phenomenology known to one type of object may have its counterpart in the other. The detection of very high energy $\gamma$-rays from the X-ray binary \ls\ by the \hess\ collaboration would appear to vindicate this conjecture \citep{science}. Observations in the 1990s have established that blazars, a subtype of AGN, emit high energy (HE, GeV) and very high energy (VHE, TeV) $\gamma$-rays \citep{fleury}. Blazars have their jet aligned close to the line-of-sight, resulting in a very strong relativistic boosting of the jet non-thermal emission. The same configuration could be found in an X-ray binary and this is perhaps the case in \ls.

\ls\ is composed of a O6.5V star and an unidentified compact object in a 3.9 day orbit \citep{motch,clark,mcswain1,mcswain04,casares}. Much of the interest in this object stems from its tentative association with the {EGRET} source 3EG J1824-1514 and from the fact that its radio emission was shown to be extended on a milliarcsecond scale \citep{paredes}. {By analogy with blazars, it is tempting to attribute the radio and $\gamma$-ray emission to high energy particles accelerated in a jet}. \ls\ is very similar to \lsi\, an X-ray binary with a B0Ve star in a 26.5~day orbit whose association with a high energy source dates back to COS-B \citep{gregory,hutchings,casares2}. The associations rely purely on positional coincidence and this is fairly poor, the {EGRET} 99\% confidence levels being roughly at 0.5\degr\ {for \ls} and 0.2\degr\ {for \lsi\ (3EG J0241+6103)}. The \hess\ detection of VHE emission coincident to within an arcmin with the position of \ls\  identifies this X-ray binary as the source of $\gamma$-rays. By extension, it places the {EGRET} tentative identification on a firmer footing.

VHE $\gamma$-ray emission is at least evidence that particles can be accelerated to very high energies in X-ray binaries. However, is it necessarily associated with blazar-like model, involving accretion of material, ejection and non-thermal acceleration in a relativistic jet? \citet{maraschi} pointed out early on that the spindown of a young, rapidly-rotating pulsar could power the $\gamma$-ray emission of \lsi. The relativistic pulsar wind would be confined by the stellar wind from the massive companion. Particles accelerated at the termination shock then produce the non-thermal emission. 

There is at least one system where such a scenario is undoubtedly at work: \psrb\ (=\object{SS 2883}), a 47.7~millisecond (ms) radio pulsar in a 3.4~yr orbit around a B2Ve star \citep{johnston94,manchester95}. Variable $\gamma$-ray emission was detected by \hess\ during periastron passage \citep{psrb}. In \ls\ and \lsi, the discovery of resolved radio emission was {interpreted as the} signature of a relativistic jet and recently, the attention seems to have focused on accretion/ejection scenarios (but see \citealt{leahy2004,martocchia}). The pulsar scenario has become unwonted. Yet, pulsars interacting with their surrounding medium can also lead to large scale emission, the bow shock nebula of high velocity ms pulsars even taking a comet tail appearance \citep[e.g.][]{gaensler}. The resolved `jets' of \ls\ and \lsi\ might be {small-scale} analogues, but with the trail originating from the interaction with a companion wind rather than with the ISM.

The similarities between \ls\ and \lsi\ have long been noticed: resolved radio emission, hard steady X-ray fluxes and {EGRET} emission in the GeV range at a level $\sim 10^{35}$~erg~s$^{-1}$. \lsi, located in the Northern hemisphere and inaccessible to \hess, has yet to be detected at TeV energies. While \psrb\ shares the property of being a TeV emitter with \ls, \lsi\ and \psrb\ both have similar periodic radio outbursts that may be linked to both of them having a Be type companion. \psrb\ has not been detected at GeV energies by {EGRET} \citep[including at periastron passage, ][]{ta96}, perhaps because its long orbit only takes it close to the star for periods of time that are too limited for the sensitivity of the instrument.

At closest, \psrb\ is about 0.7~AU away from its companion, equivalent to the apastron orbital separation in \lsi. The periastron separation of 0.1~AU in \lsi\ is itself comparable to the orbit in the more compact \ls\ (Table~\ref{orbits}). There is a scaling between the three systems that, should they be governed by the same underlying processes, might lead to interesting comparisons. {Indeed, their spectral energy distributions are similar in shape and luminosity (Fig.~\ref{sed}), suggesting the same physics might be} at work in these three known examples of ``$\gamma$-ray binaries''.

This work purports to investigate the pulsar wind scenario in light of the latest observations of these binaries. Reviewing current evidence, it is argued in \S2 that there is as yet no reason to discard this possibility in \ls\ and \lsi. The scenario is then developed in the following sections. The basic assumptions are described in \S3, showing how the physical quantities scale with the parameters of the model. Simple expectations for the spectral energy distribution are derived in \S4. A numerical model of the emission is elaborated in \S5 and applied to the $\gamma$-ray binaries in \S6. It is shown that the radio emission from the shocked pulsar wind material can look like that of a compact jet emanating from an accreting microquasar. The conclusion set out in \S7 is that the pulsar wind scenario provides a plausible, coherent and persuasive framework to interpret the emission from these $\gamma$-ray binaries.

\begin{table}
 \centering
  \caption{Orbital parameters {adopted for} the $\gamma$-ray binaries.\label{orbits}}
  \begin{tabular}{@{}lcccccc@{}}
\hline\hline
System & $P_{\rm orb}$ & $e$  & $M_2$ & $d_{S}$ & $R_\star$ & $T_\star$\\
 & (d) &  & ($M_\odot$) & (AU) & ($R_\odot$) & ($10^4$~K)\\
\hline
\ls & 3.9 & 0.35 & 23 & 0.1-0.2 & 9.3 & 3.9  \\
\lsi & 26.5 & 0.72 & 12 & 0.1-0.7 & 10 & 2.2  \\
\psrb & 1237 & 0.87 & 10 & 0.7-10 & 10 & 2.7  \\
\hline
\end{tabular}
\end{table}

\section{Accretion or rotation-powered ?}

Several observations could securely tilt the balance towards the `black hole jet', accretion-powered scenario or the `pulsar wind', rotation-powered scenario in \ls\ and \lsi\ without detailed modelling. These are examined here in turn.

\subsection{Is the compact object a black hole ?}
A compact object mass greater than 3 M$_\odot$ would rule out the pulsar wind scenario. Constraints have been set in both \ls\ and \lsi\ by measuring the radial velocity of the stellar companion. This enables the measure of the period, mass function, eccentricity and periastron angle. The mass function provides a strict lower limit to the mass of the compact object. In high-mass X-ray binaries the mass function is too small to provide a useful constraint: $f(M)=0.053$~M$_\odot$ in \ls, $f(M)=0.011$~M$_\odot$ in \lsi\ (\citealt{casares2,casares} who have the best available datasets).

A constraint on the compact object mass then requires additional knowledge on the mass of the stellar companion and the orbit inclination. Atmosphere fitting of the spectrum leads to an estimate for the mass (and radius) of the {massive star}. Accuracy requires detailed stellar models and is inherently difficult due to the high mass loss rates of early-type stars. Lines can originate at different depths and velocities in the strong (variable) stellar wind, leading to confusing measurements of radial velocities and difficulties in spectral modelling \citep[see e.g. discussion in][]{bahcall}. Parameters thereby derived have higher, probably irreducible, systematics than the quoted (statistical) errors might suggest \citep[e.g. Table~2 of ][]{casares}. In fact, the orbital period of \lsi\ (26.5~days) cannot be determined from optical spectroscopy alone {with high confidence} and radial velocity studies use as an {\em a priori} the well-determined radio outburst period \citep{hutchings,gregory02,casares2}.

The parameter space derived by \citet{casares2,casares} using these techniques allows for both a neutron star or a black hole in \ls\ and \lsi. The radial velocity amplitude implies a low orbit inclination if the compact object is massive ($i\approx 30\degr$ for $M>3 $M$_\odot$ in both \ls\ and \lsi), a good omen for a model with a relativistic jet pointed close to the line-of-sight. On the other hand, a 1.4~M$_\odot$ neutron star implies in both objects $i\approx 60\degr$. In \ls, the lack of X-ray eclipses (assuming {point-like} emission) by a companion of $\approx 9$~R$_\odot$ requires $i\la 65\degr$, consistent with the above. Line broadening of the lines gives an upper limit on their rotation speeds which, since it must be lower than breakup speed, can be used to set a lower limit on the inclination for aligned spin/orbital axis. \citet{casares2,casares} find $i>10\degr$ in both systems, again consistent with both a black hole or neutron star. Since the value of $\sin i$ is more likely to be high than low for a random distribution of inclinations, a neutron star would seem to be favoured over a black hole.

The binary orbit of \ls\ is quite compact for a high-mass X-ray binary ($P_{\rm orb}\approx 3.9$~days) and is only mildly eccentric ($e\approx 0.35$). By comparison, \lsi\ has $P_{\rm orb}\approx 26.5$~days and $e\approx0.7$.  Orbital circularization and synchronization might be relatively advanced in \ls. This has prompted \citet{casares} to use their measurement of rotational broadening {to constrain the orbit} in \ls\ by assuming that that the {massive star} is corotating at the angular orbital velocity at periastron. The compact object mass then becomes 2.7-5.0~M$_\odot$ with $i=24.9\pm2.8\degr$, probably optimistic because of the possible systematics, but in practice ruling out a neutron star. 

This conclusion is not robust. Observations of the stellar companions in other high-mass X-ray binaries with similar periods do not systematically show corotation \citep{conti}. Although synchronization is undoubtedly fast by astronomical standards (some detached early-type binaries are observed to be circularized), short-lived wind-fed HMXBs do not have ages much greater than their typical synchronization timescale. These timescales are difficult to calculate precisely: in massive stars, they depend on excited modes in the radiative envelope \citep{zahn}. The age of the system can only be guessed at. Rotation-powered emission would probably imply an energetic pulsar with a short spindown timescale. Synchronization would then not have happened. 

In both \lsi\ and \ls, optical spectroscopy alone is unable to decide between a neutron star and black hole compact object. This work will assume a 1.4~M$_\odot$ neutron star.

\begin{figure}[t!]
\resizebox{8cm}{!}{\includegraphics{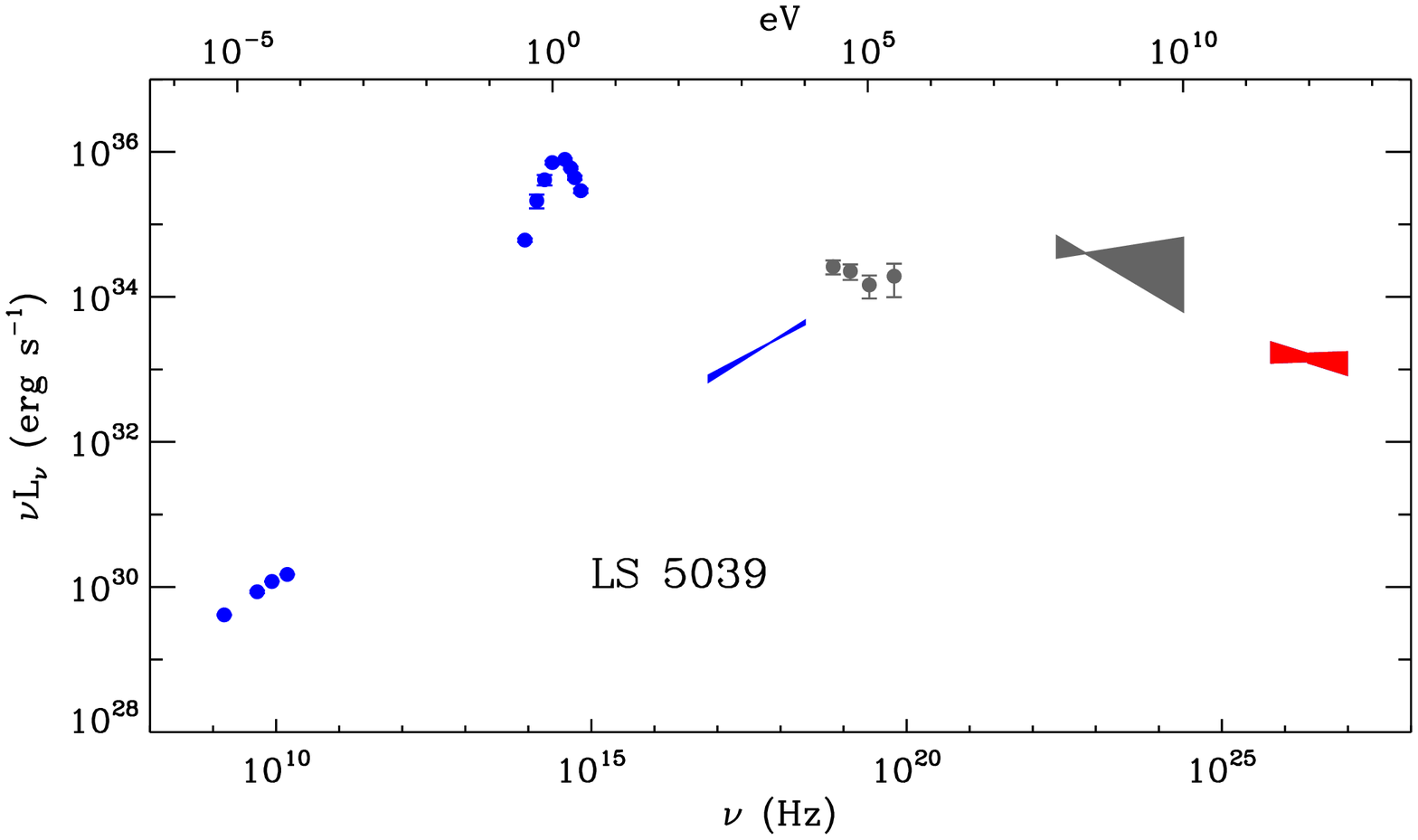}}
\resizebox{8cm}{!}{\includegraphics{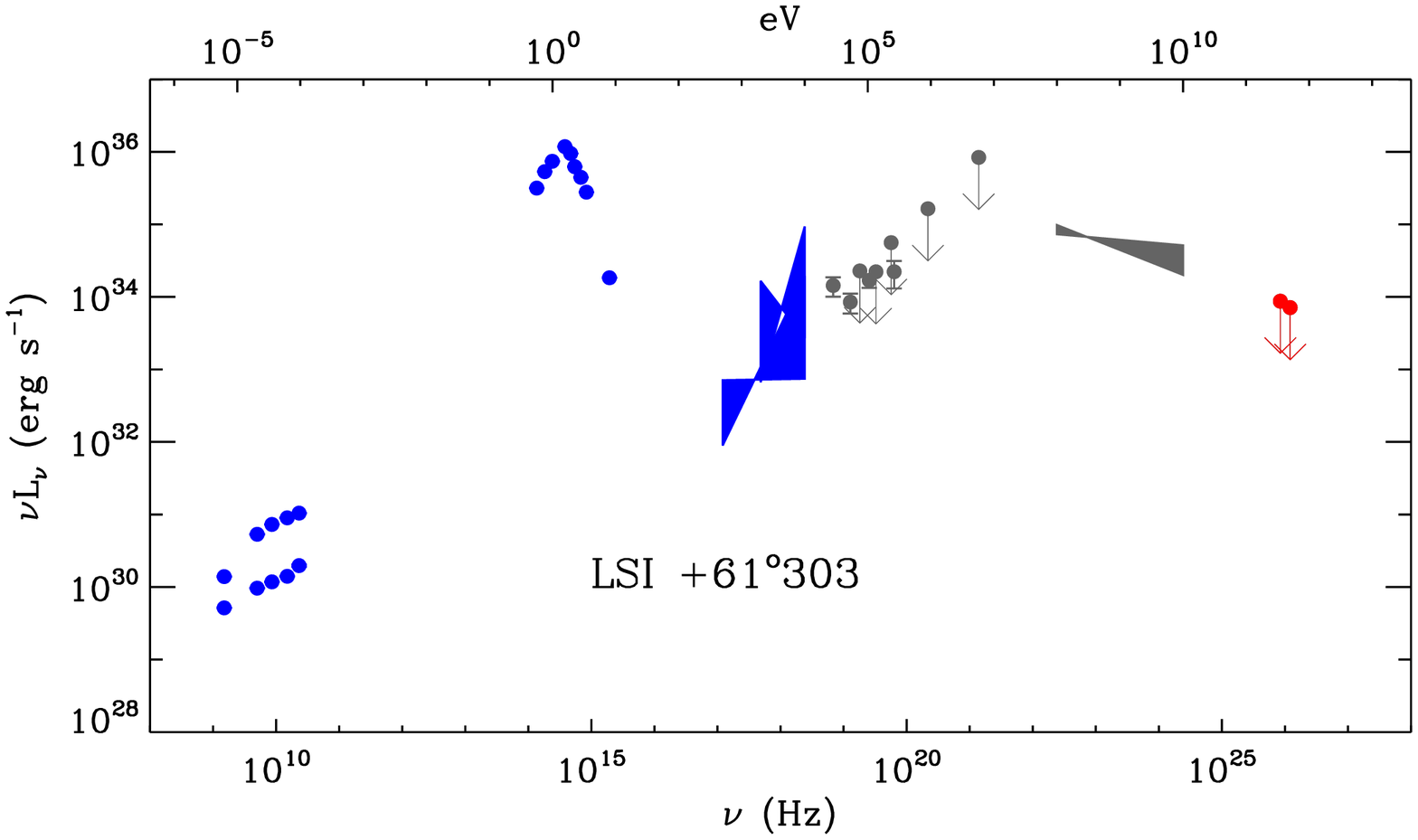}}
\resizebox{8cm}{!}{\includegraphics{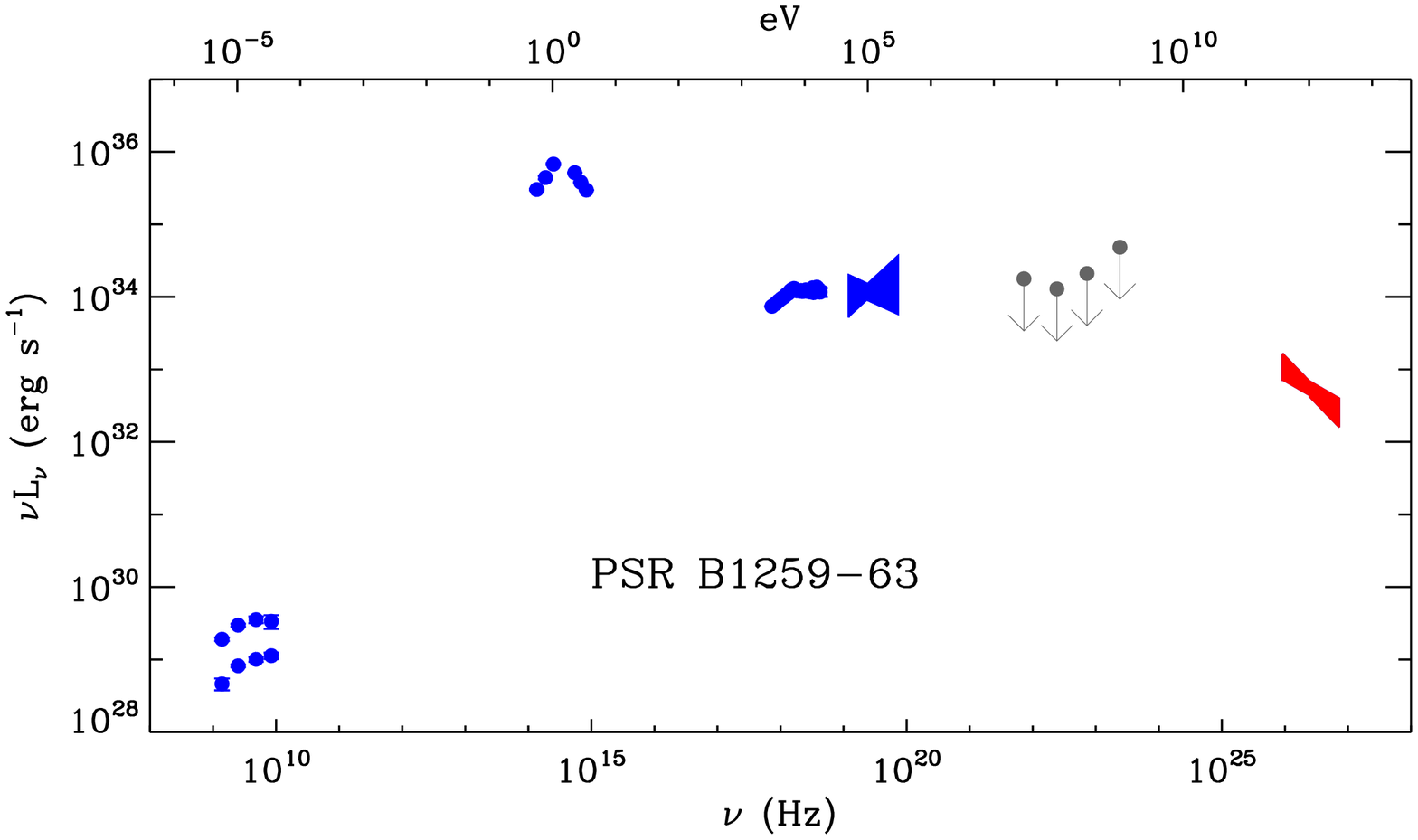}}
\caption{Observed spectral energy distributions of \ls, \lsi\ and \psrb. Fluxes have been transformed to spectral luminosities assuming distances of {2.5}~kpc (\ls, \citealt{casares}), 2.3~kpc (\lsi, \citealt{gregory79}) and 1.5~kpc (\psrb, \citealt{johnston94}). The optical-UV emission peaking at $\sim$ 1~eV in the binaries is the light from the stellar companion (uncorrected for absorption by the interstellar medium). Radio to X-ray flux values in \ls\ are taken from (in order of increasing frequency) \citet{martiLS,clark,martocchia}; in \lsi\ from \citet{strickman,Marti95,maraschi1,leahy1997,greiner}; in \psrb\ from \citet{johnston05,cutri,marraco82}. The {\em RXTE} spectrum of \psrb\ obtained during the 2004 periastron passage was kindly provided by B. Giebels. The upper envelope of radio measurements in \lsi\ and \psrb\ corresponds to values during radio outbursts. The hard X-ray {BATSE} points in \ls\ and \lsi\ are from \citet{harmon}; {+the OSSE bowtie is from \citet{grove} and the HE $\gamma$-ray {EGRET} boxes are from \citet{hartman}; {EGRET} upper limits in \psrb\ are from \citet{ta96}. The VHE $\gamma$-ray {HESS} measurements are from \citet{science,psrb} and the {Whipple} upper limits for \lsi\ }from \citet{fegan}.}
\label{sed}
\end{figure}

\subsection{Is there accretion?}

Any sign of accretion on the compact object would dismiss the pulsar wind scenario. Accretion would have to be wind-fed: Roche lobe overflow as occurs in Cyg X-1 and some other HMXBs would entail much larger {X-ray} luminosities. The orbits and sizes of the {massive stars} in \ls\ and \lsi\ support this conclusion \citep{casares2,casares}.  The rate at which mass is accreted from the stellar wind can be estimated. The Bondi capture radius is $r_a=2 GM_c/v_w^2$, where $v_w$ is the wind speed (about 2000~km~s$^{-1}$) and $M_c$ is the mass of the compact object. The Bondi mass accretion rate is then $\dot{M}\approx \dot{M}_w/ (2r_a/d_s)^2$ where $\dot{M}_w$ is the stellar wind mass loss rate {(about $10^{-7} M_\odot$~yr$^{-1}$)} and $d_s$ is the orbital separation. Using the \ls\ orbital and stellar parameters {in Table~1} gives between $0.6$--$20~10^{14}$ g~s$^{-1}$ along the orbit in \ls, {with a time-average of $5~10^{14}$ g~s$^{-1}$}. With 10\% radiative efficiency, this is {too close to the sustained} GeV emission at a level $\sim 10^{35}$~erg~s$^{-1}$. In addition, although the X-ray luminosity is comparable to those of wind-fed X-ray binaries, the lightcurve does not show evidence for the large expected associated orbital variability \citep{reig,bosch}. The slow, dense Be equatorial wind in \lsi\ {allows for higher accretion rates, in better agreement with the total luminosity and the factor $\sim 10$ variations observed in the 0.5-2~keV band lightcurve} (less above, \citealt{harrison};  \citealt{Marti95}, \citealt{paredes97}).

There are no signs of the large excursions in luminosity or spectra typical of (hard or soft) X-ray transients. The X-ray spectra are reminiscent of the low/hard X-ray state of LMXBs. There are some radio/X-ray fluctuations on timescales $\sim$ hour in both objects \citep{harrison,bosch}; but strong, fast variability on timescales of $\sim$ second, expected from a low/hard state LMXB or accreting HMXB \citep{vdk}, has not been detected \citep[][although the low flux of the sources should be noted]{ribo99,harrison}. The Thomson opacity is low in the stellar wind, hence variability might be smoothed out by scattering in a dense corona. There is no sign of such a corona in X-ray spectra. Indeed, the {apparent} lack of spectral cutoffs around either 30~keV or 100~keV {(see Fig.~1)} puts them apart from both accreting HMXBs and low state LMXBs. {\em RXTE} spectra of \ls\ did show a strong, broad Fe line \citep{ribo99}, but that could not be confirmed by other X-ray missions \citep{martocchia}. \ls\ lies in the Galactic Plane and the Fe line is very likely due to Ridge emission sampled by the large {\em RXTE} non-imaging aperture \citep{bosch}.

The radio and hard X-ray emission would be compatible with most of the emission coming from a non-thermal distribution of particles in a relativistic jet \citep{romero,paredes05,dermer2}. Gamma-rays would then be up-scattered photons. The puzzle is how such a jet could be fed by accretion without any of the usual signatures, most notably variability. In the $\gamma$-ray binaries, the broad-band SEDs are stable even on timescales of years. 

The simplest {hypothesis} is that, as in \psrb, accretion does {\em not} occur in \ls\ and \lsi. If the infall of material is stopped by pressure from a relativistic pulsar wind, this will produce very little emission and variability. Instead, the emission will be due to the shocked pulsar wind. As in the relativistic jet scenario, a non-thermal distribution of particles provides the necessary ingredient to explain the SED. The spindown of the pulsar provides a stable energy source on human timescales. The termination shock of the contained pulsar wind provides the conditions for particle acceleration. Only very mild variations are expected, due to fluctuations of the environment (stellar wind), {consistent} with observations.

\subsection{Is the radio emission a microquasar jet?}
The detection of {relativistic} motion in the radio emission, like that of the microquasar GRS 1915+105, would secure the jet model. The radio morphology of \ls\ and \lsi\ looks more like the compact jets seen in the low/hard X-ray state of low-mass X-ray binaries or Cyg~X-1, with the significant difference that their radio spectral indices are steeper than the usual flat indices measured in the self-absorbed jets. The indices ($F_\nu\sim \nu^{-0.5}$) rather hint at optically thin synchrotron emission from a $dN\propto E^{-2}dE$ power-law distribution of electrons.

The period of the radio outbursts of \lsi\ is very stable and is used as a clock for the orbital motion \citep{gregory02}. The radio luminosity of \ls\ has been steady at the same level for years \citep{marti98}. The radio brightness asymmetry of \ls\ and \lsi\ has been used to place constraints on (de)boosting of the putative relativistic jet. This yields a mild bulk flow velocity of 0.2-0.3c \citep{paredes,Massi2004}. There is no evidence in either system for highly relativistic bulk motion directed more or less towards the line-of-sight; the main argument favouring a jet interpretation is based on the morphology of the radio emission.

As mentioned in \S1, fast-moving young and old millisecond pulsars leave trails of X-ray, optical and radio emission that can reach parsec lengths, and whose properties scale with the distance where the wind terminates due to containment by the surrounding interstellar medium \citep{wang,arons}. Even the collision of standard stellar winds can produce detectable large-scale, jet-looking, radio emission \citep[e.g.][]{dougherty}. The radio emission of \ls\ and \lsi, seems to reach scales of up to several 100~AUs \citep{paredes02,Massi2004}, double-sided in the case of \ls. Energetically speaking, this emission can easily be a pulsar nebula. However, isolated ms pulsars have time to develop long (one-sided) nebula away from their direction of proper motion. In the $\gamma$-ray binaries, the direction of such a nebula would be affected on short timescales by orbital motion.  Radio observations of \lsi\ have shown curved radio emission on a scale of 50~mas, interpreted as due to precession of a jet as in SS 433 \citep{Massi2004}. Could this morphology instead be a result of the pulsar orbital motion?

In this work, it will be argued that the resolved radio emission in \ls\ and \lsi\ is {\em not} due to a relativistic jet akin to those of AGN or microquasars (and implying accretion), but arises from shocked pulsar wind material outflowing from the binary.

\subsection{Can the compact object be a pulsar? \label{pulsar}}
The X-ray and radio properties of \ls\ and \lsi\ sets them apart from other X-ray binaries, with no {conclusive} signs of accretion going on. At the same time, the luminosity, hard X-ray power law extending to high energies, weak variability and optically thin radio emission are typical for emission from young pulsars with a high spindown power \citep[e.g.][]{gs06}. Examining the pulsar wind hypothesis therefore seems justified. 

In order to avoid disc or magnetospheric accretion, the relativistic wind of the pulsar must be able to quench the infall of stellar matter attracted to it \citep{illarionov}. If all of the pulsar spindown power $\dot{E}$ is transferred to this relativistic wind, then writing the pressures gives $\dot{E}>4\dot{M} v_w c$ where $\dot{M}$ is the Bondi rate calculated above. Writing the spindown power as a function of the pulsar magnetic field $B$ and period $P$, then $P\la 230 B^{1/2}_{12} \dot{M}^{-1/4}_{15}$~ms, with $\dot{M}_{15}=\dot{M}/10^{15}$~g~s$^{-1}$. A young, ms pulsar would push out any settling accretion flow.

The young age is not problematic. Radio searches have not conclusively identified an association with a supernova remnant \citep{frail87,marti98,ribo02}; but neither have they been able to do so for \psrb, which has a measured spindown age of 3~$10^5$ years \citep{johnston}. \citet{ribo02} put an upper limit of about a Myr to the age of \ls\ by tracing back the proper motion of the system to the plane. Nitrogen enrichment in the companion's atmosphere also suggests a young age in \ls\ {(\citealt{mcswain04}; rotational-mixing could also explain the enrichment, \citealt{casares})}.

Detecting the radio pulse would nail down this scenario. Unfortunately, free-free absorption in the dense stellar winds of \ls\ and \lsi\ {would} suppress the signal. Using the Rosseland mean opacity, the radial optical depth at 1~GHz from a point at a distance $d$ from the star is $\tau\approx 3\cdot 10^4 \dot{M}_{\rm w, 7}^2 v^{-2}_{w,2000} T_4^{-3/2} \nu^{-2}_{\rm GHz} d_{0.1}^{-3}$ for a $10^{-7}$ M$_\odot$~yr$^{-1}$  coasting wind with a speed 2000~km~s$^{-1}$ and temperature $10^4$~K; $d$ is taken to be 0.1~AU, about the periastron orbital separation in \ls\ and \lsi. High opacities are also found in \lsi\ \citep{taylor82}. In \psrb, the suppression of pulsed emission is observed close to periastron \citep{johnston,melatos}. 

Pulsed emission in the {X-ray/GeV} band would serve just as well. Such a detection does not seem realistic without prior knowledge on the pulse period derivatives given the sensitivities achievable even by {\em GLAST}. In addition, the pulsed {GeV} signal in \ls\ may be smoothed out by emission from the $e^+e^-$ cascade initiated by absorption of TeV $\gamma$-rays on stellar photons \citep{dubus}.

As the pulsar rotation gradually decreases, the $\gamma$-ray binaries {would} become standard accretion-powered high mass X-ray binaries, perhaps brighter in the X-ray sky {but probably without significant emission} at $\gamma$-ray energies. That three examples of such short-lived phases may have been found {perhaps} implies a large population of descendant HMXBs of which some might be the enshrouded hard X-ray sources discovered by {\em INTEGRAL}. Some of the unidentified {EGRET} sources located in the Galactic Plane may also turn out to be pulsars in binaries. Massive X-ray binary population synthesis calculations by \citet{meurs89} predict about 30 binaries in the brief young pulsar stage in the Galaxy, consistent with three examples being found within 3~kpc.

The conclusion is that \ls\ and \lsi\ could very well harbour a young pulsar with a strong relativistic wind like that of \psrb. A model based on this scenario is explored in the next sections.

\section{Scaling relationships in the binary plerion}

The scenario, identical to that proposed for \psrb, is the following: a young pulsar with a spindown rate of order $10^{36}$~erg~s$^{-1}$ generates a strong relativistic wind beyond the light cylinder \citep{rees}. The pulsar wind is assumed to be radial, isotropic and composed of principally mono-energetic electrons/positrons with a Lorentz factor $\gamma_p$. All the spindown power is transferred to the leptons, with a small fraction going to the magnetic field. The ratio of magnetic to kinetic energy $\sigma$ is a parameter of the model.

The pulsar wind is contained by the stellar wind (in a plerion like the Crab, the container is the supernova remnant material). A collisionless shock forms beyond which the leptons from the pulsar wind are accelerated and isotropized. The shocked pulsar wind material is separated from the stellar wind by a contact discontinuity across which mixing might occur (neglected here). The shock has a `bow' or `comet' shape with a tail extending away from the stellar companion. The radio emission occurs in this nebula. Studying such a complex interaction is simplified by noting everything scales with the standoff distance \citep[see e.g.][]{bucciantini}. The standoff distance $R_s$ is the radial distance from the pulsar to the star at which the ram pressures equilibrate (\S3.2).

Most emission occurs in the vicinity of $R_s$, whose value can be calculated given the orbital parameters, {and the stellar wind density and velocity}. Straightforward estimates indicate that the physical conditions at the shock are set by only three quantities: the shock distance $R_s$ from the pulsar, the orbital separation $d_s$ and the combination $\sigma\dot{E}$, which describes the pulsar wind energy and magnetization. The overall properties of the resulting spectral energy distribution are then deduced in the next section (\S4). A quantitative description including emission from the nebula is given in \S5-6.

\subsection{The pulsar wind}
The pulsar wind is described by the Lorentz factor $\gamma_p$ of the electrons/positrons, the spindown power $\dot{E}$ and the ratio of magnetic to kinetic energy $\sigma$. The ({proper frame}) density $n_1$ and ({observer frame}) magnetic field $B_1$ in the pulsar wind upstream of the shock are then derived from \citep{kc2}
\begin{eqnarray}
\sigma&=&\frac{B^2_1}{4\pi n_1 \gamma_p^2 m_e c^2}\\
\frac{\dot{E}}{4\pi R_s^2 c}&=&\frac{B_1^2}{4\pi} \left(\frac{1+\sigma}{\sigma}\right)\label{edot}
\end{eqnarray}
Most of the energy could actually be carried by ions, which are probably a necessary ingredient to accelerate the leptons at the termination shock \citep{arons2}. Here, the kinetic energy of the wind is {assumed} to be only in the electrons/positrons.

In \psrb\, $\dot{E}$ is $8\cdot 10^{35}$ erg~s$^{-1}$ and the pulsar magnetic field $B$ is about $3\cdot10^{11}$~G (corresponding to a spin period $P=47.7$~ms and period derivative $\dot{P}=2.3\cdot 10^{-15}$). Thereafter, $\dot{E}$ will be expected to be of order $10^{36}$~erg~s$^{-1}$. Models of the Crab plerion yield values of order $10^{-3}$ for $\sigma$ and $\gamma_p$ in the $10^5-10^6$ range \citep{rees,kc1}. Explaining the implied conversion of electromagnetic energy (close to the pulsar) to kinetic energy (far from the pulsar) is a key question of pulsar physics. Values for $\sigma$ could be higher if the shock occurs closer to the pulsar than in Crab, as proposed for \psrb\ by \citet{ta97}; in any case $\sigma$ will be expected smaller than 1. 

\subsection{The standoff distance}
The standoff distance point $R_s$ of the pulsar wind is given by
\begin{equation}
\frac{\dot{E}}{4\pi R_s^2 c}=\rho_w (\mathbf{v}_w-\mathbf{v}_p)^2
\label{standoff}
\end{equation}
where $\mathbf{v}_p$ is the pulsar radial orbital speed calculated from the binary parameters, $\rho_w$ is the stellar wind density and $\mathbf{v}_w$ the stellar wind speed at the standoff distance. An estimate of $R_s$ can be obtained {by balancing the pressures of the two flows}, assuming a stellar wind coasting at a speed $v_\infty$ and a negligible orbital speed:
\begin{equation}
R_s\approx \frac{d_s}{1+(P_w /\dot{E})^{1/2}}\label{full}
\end{equation}
where $P_w/c=\dot{M}_w v_\infty$ is the stellar wind momentum flux with $\dot{M}_w$ the stellar mass loss rate, supposed isotropic here, and {$d_s$ is the orbital separation}. To fix ideas, the radiatively-driven wind of \ls\ has $\dot{M}_w\approx 10^{-7} M_\odot~\mathrm{yr}^{-1}$ and $v_\infty\approx 2000$~km~s$^{-1}$ which gives $P_w \approx 5\cdot 10^{37}$~erg~s$^{-1}$ \citep{mcswain04}. When the stellar wind is more powerful than the pulsar wind, in the sense that $P_w\gg \dot{E}$, then the standoff point can be written as 
\begin{equation}
R_s\approx 1.5\cdot10^{11}\mathrm{~cm~} d_{0.1} P_{w,38}^{-1/2}\dot{E}_{36}^{1/2}
\label{rs}
\end{equation}
where $d_{0.1}=d_s/0.1~\mathrm{AU}$ is the orbital separation (0.1~AU at periastron in \ls\ and \lsi), $\dot{E}_{36}=\dot{E}/10^{36}$~erg~s$^{-1}$ and $P_{w,38}=\dot{M}_w v_\infty c/10^{38}$~erg~s$^{-1}$. The standoff distance stays at about a tenth of the binary separation throughout the orbit. Writing down explicitly the dependence on the stellar wind (Eq.~\ref{rs}) is easy using $v_\infty$ and $P_w$. Eq.~\ref{full} is used for the slow equatorial winds of Be stars (\S6). In the following, it proved more convenient to explicitely keep $R_s$ as a parameter, {derived under the plausible assumptions described above. Some of the uncertainties in the geometry of the shock region or the stellar wind parameters (particularly in the wind acceleration region) are thereby hidden in $R_s$.}

\subsection{Downstream flow at standoff}
Downstream values for the magnetic field and particle density are calculated using perpendicular MHD shock jump conditions at $R_s$ \citep{kc1}. The post-shock magnetic field is (for $\sigma\ll1$)
\begin{equation}
B= 3(1-4\sigma) \left(\frac{\dot{E}/c}{R_s^2}\frac{\sigma}{1+\sigma}\right)^{1/2}\approx 5~ (\dot{E}_{36}\sigma_3)^{1/2} R_{11}^{-1}~\mathrm{G}
\label{magnetic}
\end{equation}
where $\sigma_{3}=\sigma/10^{-3}$ and $R_{11}=R_s/10^{11}$~cm. 

The downstream particles are assumed to be accelerated to a canonical $dn_\gamma \propto \gamma^{-2} d\gamma$ power law from a minimum energy $\gamma_{\rm min}$ to a maximum energy $\gamma_{\rm max}$. For a given $\gamma_{\rm max}$, the normalisation and lower limit $\gamma_{\rm min}$ of the post-shock particle distribution are found by matching to the downstream particle and energy densities. For a $\gamma^{-2}$ power law and $\sigma\ll 1$, $\gamma_{\rm min}$ is given by \citep{kc2}
\begin{equation}
\gamma_{\rm min} \ln \left(\frac{\gamma_{\rm max}}{\gamma_{\rm min}}\right) \approx \frac{\gamma_p}{\sqrt{2}}
\label{gammin}
\end{equation}
For a post-shock distribution covering 2-3 decades in energy, $\gamma_{\rm min}\approx 0.1 \gamma_p$.

\subsection{Timescales at standoff\label{timescales}}
Particles are assumed to be accelerated efficiently up to the maximum energy $\gamma_{\rm max}$ {above} which the timescale at which particles radiate their energy becomes shorter than the acceleration timescale. The gyroradius is used to give an estimate of the acceleration timescale to a Lorentz factor $\gamma$
\begin{equation}
t_{\rm acc}\approx\frac{\gamma m_e c}{e B}\approx 0.01~ \gamma_6 ~(\dot{E}_{36} \sigma_3)^{-1/2} R_{11}~\mathrm{seconds}
\end{equation}
using the downstream magnetic field (Eq.~\ref{magnetic}) and writing $\gamma_6=\gamma/10^6$. For ion dominated shocks \citep{arons2}, the relevant timescale is given by the ion gyroradius and the maximum electron energy is $\sim \gamma_p m_p/m_e $ ($\gamma_p$ is then the upstream ion Lorentz factor). The maximum $\gamma$ reached must have a gyroradius smaller than the characteristic scale of the shock $R_s$, implying $\gamma< 3\cdot 10^8 ~(\dot{E}_{36}\sigma_3)^{1/2}$. {Particles do not reach such high $\gamma$ as they then lose energy to radiation more rapidly than they can be accelerated}. 

The synchrotron timescale is 
\begin{equation}
t_{\rm sync}=\frac{9}{4}\frac{m_e^3 c^5}{e^4\gamma}\frac{1}{B^2}\approx 30~ \gamma_6^{-1}~(\dot{E}_{36}\sigma_3)^{-1} R_{11}^{2}~\mathrm{seconds} 
\label{tsync}
\end{equation}
so the maximum $\gamma$ such that $t_{\rm sync}>t_{\rm acc}$ is
\begin{equation}
\gamma_{\rm max}\approx5~10^{7}~(\dot{E}_{36}\sigma_3)^{-1/4}~R_{11}^{1/2}
\label{gammax}
\end{equation}
which, in principle, could result in a maximum photon energy of 25~TeV. Photons up to energies $\ga 4$~TeV are detected by \hess\ \citep{science}, requiring electrons with Lorentz factors high enough that $\gamma_{\rm max} m_e c^2 \ga$ 4~TeV, i.e. $\gamma_{\rm max}\ga 10^7$. This is below the limit from synchrotron losses (Eq.~\ref{gammax}) for all plausible values of the pulsar $\sigma$-parameter.

\begin{figure}
\resizebox{7cm}{!}{\includegraphics{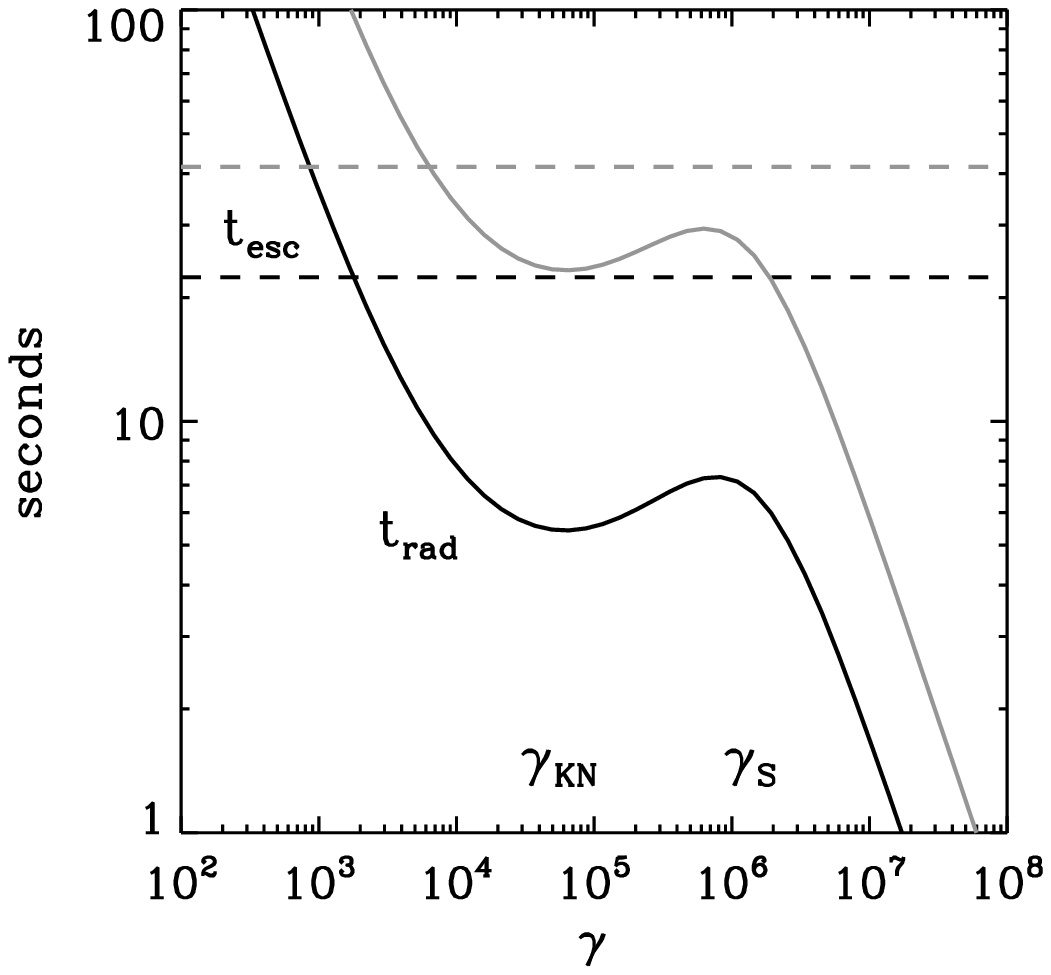}}
\caption{Timescales at the standoff point as a function of the electron Lorentz factor $\gamma$ in \ls. The radiative timescale $t_{\rm rad}^{-1}=t_{\rm sync}^{-1}+t_{\rm IC}^{-1}$ (solid lines) and the escape timescale $t_{\rm esc}$ (dashed lines) are shown at periastron (black) and apastron (grey). Synchrotron cooling dominates above $\gamma_{\rm S}\approx 2 \cdot 10^6$. Inverse Compton cooling dominates below, the transition between Thomson and Klein-Nishina regimes occurring at $\gamma_{\rm KN}\approx 6\cdot10^4$. $R_s$ is $2 \cdot 10^{11}$~cm at periastron ($d=0.1$~AU) and $4\cdot 10^{11}$~cm at apastron ($d=0.2$~AU) (see Table~\ref{freq}).}
\label{time}
\end{figure}

The inverse Compton timescale in the Klein-Nishina regime $\gamma h\nu\gg m_e c^2$ is \citep{bg}
\begin{equation}
t_{\rm KN}\approx5.2~\gamma_6 \left(\frac{3~\mathrm{eV}}{kT_\star}\right)^2 \left(\frac{d_s}{R_\star}\right)^2 \left[\ln \left(\frac{4 \gamma k T_\star}{m_e c^2}\right) -1.98 \right]^{-1}~\mathrm{seconds}
\end{equation}
where the stellar (blackbody) radiation density has been lowered by geometric dilution and $R_s$ is assumed to be small compared to the orbital separation $d_s$. The stellar temperature $T_\star$ varies between 2--4$\cdot10^4$~K with {the stellar radius} $R_\star\approx10R_\odot$ for \ls, \lsi\ and \psrb\ so that
\begin{equation}
t_{\rm KN}=20~\gamma_6 d_{0.1}^2 \left[\ln \gamma_6 +1.3 \right]^{-1}~T_{\star,4}^{-2} R_{\star,10}^{-2}\mathrm{seconds}
\label{tkn}
\end{equation}
with $T_{\star,4}=T_\star/40,000$~K and $R_{\star, 10}=R_\star/10 R_\odot$. In the Thomson limit $\gamma h\nu\ll m_e c^2$ then
\begin{equation}
t_{\rm T}=\frac{9}{4}\frac{m_e^3 c^5}{e^4\gamma}\frac{1}{8\pi U_\star}\approx
40~\gamma_3^{-1}d_{0.1}^{2}T_{\star,4}^{-4}R_{\star,10}^{-2}~\mathrm{seconds}
\label{thomson}
\end{equation}
where $U_\star=(\sigma_B T_\star^4/c) (R_\star/d_s)^2$ is the stellar photon density. Most of the stellar photons have an energy of 2.7$k T_\star\approx 9$~eV so that the turning point between Thomson and Klein-Nishina regime is for  $\gamma_{\rm KN}\approx$~511~keV/9~eV$\approx 6\cdot10^4$. The interaction timescale is then at a minimum of $\approx 7$~s (see Fig.~\ref{time}). 

The timescale for particles in the pulsar wind to reach the standoff point is
\begin{equation}
t_w\approx R_s/c\approx 3~R_{11}~\mathrm{seconds}
\label{tw}
\end{equation}
which is only slightly shorter than the minimum inverse Compton timescale derived above. This is important since if the pulsar wind is composed of mono-energetic particles with $\gamma_p\approx 6 \cdot10^4$, then a fraction of the spindown power can be radiated by bulk inverse Compton interactions in $\gamma$-rays before reaching the termination point (Compton drag). There may then not even be a shock. (The upstream magnetic field is frozen in the MHD flow so upstream particles do not radiate synchrotron.) The resulting $\gamma$-ray spectrum would display line-like emission with a strong orbital phase dependence \citep{Bogovalov2000,Ball2000}. This possibility {appears} excluded by the power-law \hess\ spectra extending to a few~TeV in \ls\ and \psrb\ \citep{science,psrb}. Compton losses are weaker in \lsi\ and \psrb\ (smaller $T_\star$ and larger separations) than in \ls.

Downstream of the shock, the inverse Compton timescale is greater than the synchrotron timescale when $\gamma<\gamma_{\rm S}$ where
\begin{equation}
\gamma_\mathrm{S} \approx 1.2\cdot10^6 ~(R_{11}/d_{0.1}) (\dot{E}_{36}\sigma_3)^{-1/2} T_{\star,4} R_{\star,10}
\label{gammas}
\end{equation}
using Eq.~\ref{tsync} and \ref{tkn}. The maximum electron energy $\gamma_{\rm max}$ is indeed set by synchrotron losses (Eq.~\ref{gammax}), and not by inverse Compton losses. Perhaps counter-intuitively, high energy particles lose their energy mostly through synchrotron radiation, although the radiation density at $R_s$ from the luminous stellar companion is much larger than the magnetic field energy density ($U_\star/U_B\approx 10^4 (R_{11}/d_{0.1})^2(\dot{E}_{36}\sigma_3)^{-1}$). The difference in densities is compensated by the reduction of the inverse Compton cross-section in the Klein-Nishina regime.  Note that $\gamma_\mathrm{S}$ is independent of the orbital separation when the stellar wind dominates ($R_s\propto d_s$, see Eq.~\ref{rs}).

When $\sigma\ll 1$, the post-shock material is found from the jump conditions to initially flow away at $v \approx c/3$ \citep{kc1} so the escape timescale is of order
\begin{equation}
t_{\rm esc}\approx\frac{R_s}{c/3}\approx 10~ R_{11}~\mathrm{seconds}
\end{equation}
Here, $\gamma_{\rm min}\approx 0.1\gamma_p>\gamma_{\rm KN}$ so particles radiate in the Klein-Nishina regime up to $\gamma_{\rm S}$. Their radiative timescale $t_{\rm IC}\approx$ 7--20~s is comparable to their escape timescale. Particles with $\gamma>\gamma_\mathrm{S}$ lose energy to synchrotron radiation, on a timescale $t_{\rm sync}\la 30$~s that is, again, comparable to their escape timescale. The synchrotron radiative efficiency increases with decreasing $R_s$ since $t_{\rm sync}/t_{\rm esc}\sim R_s$.

\section{Expected high energy spectrum}

The timescales given above are sufficient to get an approximate idea of the expected high energy emission from the pulsar wind shock region. The aim is to provide a simple framework to understand the results from the more complex modelling presented in \S5-6.

\subsection{Spectral shape \label{spectral}}
Since the radiation timescales are comparable to or (much) smaller than the flow timescale, the situation can be approximated to steady-state injection of power-law electrons interacting in the Klein-Nishina regime with an isotropic bath of quasi mono-energetic photons. \citet{moderski} have studied exactly this case, taking into account synchrotron emission in a uniform magnetic field. The overall characteristics of the spectrum emitted in the vicinity of the pulsar can be deduced from their results and the calculations above.  If particle cooling is not fast enough, then particles will travel to regions of lower energy densities, modifying the spectral energy distribution from that presented here. The steady state assumption is reasonable for the high energy spectrum (\S5).

\citet{moderski} show that the steady-state electron distribution displays three regimes: (a) at low energies, interactions in the Thomson limit steepen, as usual, the distribution by 1 in index; (b) at high energies, synchrotron emission also steepens the distribution by 1; (c) at intermediate energies, where inverse Compton interactions in the Klein-Nishina regime dominate, the decrease in cross-section is such that the distribution index is unchanged or hardened at most by $\la 0.5$. The hardening decreases with the ratio $U_\star/U_B$. The change between regimes (a) and (b) occurs at the transition to the Klein-Nishina regime $\gamma_{\rm KN}$. Case (a) does not happen here since we initially have $\gamma_{\rm min}> \gamma_{\rm KN}$. The change between regimes (b) and (c) occurs at $\gamma_\mathrm{S}$ where synchrotron losses start to dominate. A $dN\sim \gamma^{-2}d\gamma$ injection spectrum will therefore lead to a steady-state electron distribution going from $dN\sim \gamma^{-1.5} d\gamma$ to $\gamma^{-3}$ above $\gamma_{\rm S}$.

Using the downstream magnetic field (Eq.~\ref{magnetic}), the characteristic synchrotron frequency for an electron with an energy $\gamma m_e c^2$ is
\begin{equation}
h\nu \approx 100~\gamma_6^2 (\dot{E}_{36}\sigma_3)^{1/2} R_{11}^{-1}~~\mathrm{keV}
\end{equation}
The three frequencies relevant to the spectral shape, corresponding to $\gamma_\mathrm{min}$, $\gamma_{\rm S}$ and $\gamma_{\rm max}$ are (using $\gamma_{\rm min}\approx 0.1\gamma_p$ and Eqs.~\ref{gammax},\ref{gammas}):
\begin{eqnarray}
h\nu_{\rm sync, min}&\approx& 1~\gamma_{p, 6}^2 (\dot{E}_{36}\sigma_3)^{1/2} R_{11}^{-1}~~\mathrm{keV}\\
h\nu_{\rm sync, S}&\approx& 150 ~(\dot{E}_{36}\sigma_3)^{-1/2}  (T_{\star,4} R_{\star,10})^2 (R_{11}/d^2_{0.1})~~\mathrm{keV}\label{syncs}\\
h\nu_{\rm sync, max}&\approx& 250~\mathrm{MeV}
\end{eqnarray}
The expected synchrotron spectrum \citep{moderski} is a hard power law $\nu F_\nu \sim \nu^{0.7}$ from about 1~keV initially ($\gamma_{\rm min}$) up to a break energy $\approx 100$~keV ($\gamma_\mathrm{S}$). The spectrum then flattens, with $\nu F_\nu \sim$ constant up to $\approx$ 250~MeV ($\gamma_{\rm max}$). {Since $R_s/d_s$ is almost constant (Eq.~5)}, the break frequency changes as $d_s^{-1}$ along the orbit because of the variation of the magnetic field at the standoff point. The maximum synchrotron frequency is independent of all parameters.

Inverse Compton interactions occur in the Klein-Nishina regime so that the characteristic frequencies are given by $h \nu \approx \gamma m_e c^2$
\begin{eqnarray}
h\nu_{\rm comp, min}&\approx& 50~\gamma_{p, 6}~\mathrm{GeV}\\
h\nu_{\rm comp, S}&\approx& 0.6~(\dot{E}_{36}\sigma_3)^{-1/2}(T_{\star,4} R_{\star,10}) (R_{11}/d_{0.1})~~\mathrm{TeV}\label{comps}\\
h\nu_{\rm comp, max}&\approx& 25~(\dot{E}_{36}\sigma_3)^{-1/4} R_{11}^{1/2}~~\mathrm{TeV}
\end{eqnarray}
\citet{moderski} find the inverse Compton spectrum for a $\gamma^{-2}$ injection is constant or rising slightly in $\nu F_\nu$ up to $\nu_{comp, S}$ (their Fig.~6). Above an energy $\gamma_\mathrm{S} m_e c^2$, the electron energy distribution steepens to $\gamma^{-3}$. The spectral index steepens by $\approx 1.5$ (their Eq.~31) and $\nu F_\nu\sim \nu^{-1.5}$ is expected above 0.6~TeV. 

Note that photons emitted with energies above about 30~GeV can interact with stellar photons in the binary system to produce pairs $\gamma\gamma\rightarrow e^+e^-$ \citep{gs}. This leads at least to a hardening (softening) of the observed VHE $\gamma$-ray spectrum above (below) a few 100~GeV \citep{dubus}. This can produce modifications to the electron distribution since pairs are created (this is further discussed in \S6.1.1).

\subsection{Luminosity}
Assuming the situation is steady-state and that the pulsar spindown power is largely given to the particles ($\sigma\ll 1$), the synchrotron and inverse Compton luminosities should be, to order unity, equal to $\dot{E}$. With a $\gamma^{-2}$ injection, the inverse Compton and synchrotron luminosities are roughly equal \citep{moderski}.

With such short radiative timescales, only a moderate spindown power of $10^{36}$~erg~s$^{-1}$ is sufficient to achieve a high radiative efficiency in X-rays and $\gamma$-rays. This has little to do with the assumptions made on the pulsar wind $\gamma_p$, $\sigma$ or $\dot{E}$ ({if the wind is purely composed of electrons and positrons: a pulsar wind with a significant ion fraction would require a higher $\dot{E}$ to obtain the same luminosity)}. The efficiency comes from the high radiation and magnetic energy densities at the standoff point which allow fast cooling of all of the injected particles. This contrasts with pulsars interacting with their supernova remnant or the interstellar medium, whose termination shock occurs much farther out, at correspondingly lower energy densities.

The main factor setting the high energy luminosity is the shock distance $R_s$ or, alternatively, the orbital separation $d_s$ (since $R_s\propto d_s$ in Eq.~\ref{rs}). The shock distance enters as $R_s^2$ in the synchrotron timescale compared to $R_s$ in the escape timescale. The smaller $R_s$ is, the more synchrotron emission dominates and the more the pulsar can be expected to be radiatively efficient. However, this is at the expense of the inverse Compton flux since the associated timescale stays constant with $R_s$.

The dominant radiation process does not change along the orbit for a given electron energy (because $\gamma_{\rm S}\sim R_s/d_s$), but the probability that this electron can escape from the region of high $U_\star$ and $U_B$ without significant radiation increases with $d_s$ (or $R_s$). The steady-state assumption then breaks down. Conditions change as particles flow away from the shock region. The overall nebula must then be taken into account to model the emission (\S5).

\subsection{Comparison to SEDs\label{simplesed}}
\begin{table*}
 \centering
 \begin{minipage}{115mm}
  \caption{Characteristic frequencies of the expected spectral energy distributions. The standoff distance (first number) is calculated at periastron and apastron using Eq.~\ref{standoff} and the stellar wind parameters in \S6. Both a polar and an equatorial wind are considered for the Be stars. The break frequencies in the synchrotron and Compton spectra are then calculated using Eq.~\ref{syncs} and Eq.~\ref{comps} in \S\ref{spectral}. The pulsar wind parameters are the same for all the objects: $\dot{E}=10^{36}$~erg~s$^{-1}$, $\sigma=0.01$ and $\gamma_p=10^5$. The star temperatures are such that inverse Compton interactions are in the Klein-Nishina regime. \label{freq}}
  \begin{tabular}{@{}lrrrrrr@{}}
\hline\hline
\ls & \multicolumn{3}{c}{periastron $d_s=0.1$~AU}	& \multicolumn{3}{c}{apastron $d_s=0.2$~AU} \\
\hline
& $2\cdot10^{11}$~cm & 100~keV & 0.4 TeV & $4\cdot 10^{11}$~cm & 200~keV & 0.4~TeV\\
& \\
\lsi & \multicolumn{3}{c}{periastron $d_s=0.1$~AU}	& \multicolumn{3}{c}{apastron $d_s=0.7$~AU} \\
\hline
polar	& $8\cdot10^{11}$~cm & 100~keV & 0.8 TeV & $4\cdot 10^{12}$~cm & 15~keV & 0.5~TeV \\
equatorial	& $5\cdot10^{10}$~cm & 6~keV & 0.05~TeV & $2\cdot 10^{12}$~cm & 5~keV & 0.3~TeV\\
&\\
\psrb	& \multicolumn{3}{c}{periastron $d_s=0.7$~AU}	& \multicolumn{3}{c}{apastron $d_s=10$~AU} \\
\hline
polar 	& $3\cdot10^{12}$~cm & 12~keV & 0.5~TeV & $4\cdot 10^{13}$~cm & 1~keV & 0.5~TeV \\
equatorial	& $10^{12}$~cm & 4~keV & 0.2~TeV & \multicolumn{3}{c}{not applicable}\\
\hline
\end{tabular}
\end{minipage}
\end{table*}

The characteristic frequencies for all three $\gamma$-ray binaries were calculated using the parameters in Table~\ref{orbits} and additional assumptions on their stellar winds. For \ls, the wind is radiative with  $\dot{M}_w= 10^{-7}~M_\odot$~yr$^{-1}$ and $v_\infty= 2400$~km~s$^{-1}$. For \lsi\ and \psrb\ two winds were assumed: a dense, equatorial outflow and a fast, polar wind similar to that in \ls. Details are given in \S6. The standoff distances $R_s$ and frequencies are given in Table~\ref{freq}. 

Anticipating on \S6, the pulsar was assumed to have $\gamma_p=10^5$, $\sigma=10^{-2}$ and $\dot{E}=10^{36}$~erg~s$^{-1}$ in all three systems. Note that $\dot{E}$ and $\sigma$ always appear in the same proportion so that for given $R_s$ there are really only two parameters, $\gamma_p$ and $\sigma \dot{E}$.  The characteristic frequencies are not very sensitive to $\sigma\dot{E}$.  Interestingly, the SED is also indifferent to the exact value of $\gamma_p$ but for the low energy end of the inverse Compton spectrum.

The expected breaks match reasonably well the observed SEDs of \ls, \lsi\ and \psrb, notably the rising then flattening X-ray to $\gamma$-ray spectra. An important observable feature is that, all other parameters being equal, the break frequency in X-rays $\sim 1/d_s$ should be lower at apastron than at periastron. At higher energies, the break $\nu_{\rm comp, S}$ in \lsi\ and \psrb\ occurs at frequencies a factor 2 lower than in \ls, because of the lower $T_\star$ (Table \ref{orbits}). The TeV spectrum of \psrb\ is indeed steeper than that of \ls. Unlike what happens with synchrotron, the spectral break in the inverse Compton emission should not change along the orbit if $R_s$ {is proportional to} $d_s$ (Eq.~5).

However, there are two shortcomings: the radio emission and the {EGRET} detections of \ls\ and \lsi\ at GeV energies (Fig.~\ref{sed}). The radio emission cannot be addressed by the present scheme: the minimum synchrotron frequency is $\approx$1~keV and radiation {well below must originate from particles that have cooled and escaped} from the vicinity of the pulsar (see \S5-6).

The GeV emission may be problematic as the model predicts a gap between 250~MeV and 50~GeV when $\gamma_p=10^6$. There is no room for increasing the synchrotron emission so as to include at least a fraction of the {EGRET} band: the maximum synchrotron frequency is fixed by the assumptions on $\gamma_{\rm max}$. Advocating higher $\gamma_{\rm max}$ seems unreasonable.  Moreover, for large orbital separations as in \psrb\ or \lsi\ (at apastron), the maximum electron energy is limited by the Larmor radius (supposed $\la R_s$) rather than by radiative losses. These range from 2~MeV (apastron, polar wind) to 80 MeV (periastron, equatorial wind) for \psrb\ and from 30~MeV (apastron, polar wind) to 250~MeV (periastron, equatorial wind) for \lsi. The easiest solution consists in lowering the Lorentz factor $\gamma_p$ of the particles in the pulsar wind to $10^5$. This lowers $\nu_{\rm comp, min}$ to a few GeV whilst leaving the spectral breaks unchanged. The inverse Compton timescale becomes long compared to $t_{\rm esc}$ at the low energy end of the electron distribution and, as for radio, high energy emission between 250~MeV and a GeV will also come from escaping particles.

\section{A numerical model of the pulsar nebula}
The detailed spectral energy distribution, including evolution of the physical conditions in the flow, is calculated here with the aim of accounting for the resolved radio emission as well as providing more quantitative comparisons. The assumptions are described in this section, with \ls\ used as an illustrative example when necessary. The application and comparison with observations are made in \S6. 

\subsection{Evolution of the particle distribution}
Particles are continuously injected with a $\gamma^{-2}$ power-law energy spectrum between $\gamma_{\rm min}$ and $\gamma_{\rm max}$ beyond the termination shock, in an emitting region of characteristic size $R_s$. {Other distributions have not been considered for the sake of simplicity.} The downstream particle density, in the limit of small $\sigma$ is given by \citep{kc2}
\begin{equation}
n_2\approx{3 n_1 \gamma_p}\approx 980~ \gamma_6^{-1}\dot{E}_{36} R_{11}^{-2} \mathrm{~~particles~cm}^{-3}
\end{equation}
using Eq.~\ref{edot}. The normalisation of the injected particle power-law $Q(\gamma)$ is then found from
\begin{equation}
\int_{\gamma_{\rm min}}^{\gamma_{\rm max}}Q(\gamma) d\gamma\approx \frac{c}{3} n_2\pi R_s^2\approx 3\cdot10^{35} ~ \gamma_6^{-1} \dot{E}_{36} \mathrm{~particles}~\mathrm{s}^{-1} 
\end{equation}

Particles move downstream away from the pulsar with a speed $v$ (initially $\approx c/3$) whilst emitting radiation. The resulting `cometary' nebula is modelled by dividing it into small (logarithmic) subsections where the physical parameters (magnetic field, density etc) are homogeneous. The nebula is assumed to expand conically. The evolution of the distribution function $N(\gamma,z)$ along the nebula is given by
\begin{equation}
\frac{\partial N}{\partial z}=\frac{\partial}{\partial \gamma}\left[ \frac{\mathrm{d}\gamma}{\mathrm{d} z} N  \right]+Q(\gamma)\delta(z-1)
\label{eqevol}
\end{equation}
where $z$ is the distance along the nebula normalised to the standoff distance $R_s$. The radiative and adiabatic losses for an electron (positron) are:
\begin{equation}
\frac{\mathrm{d}\gamma}{\mathrm{d} z}=
\frac{R_s}{v} \left[\left(\frac{\mathrm{d} \gamma}{\mathrm{d} t}\right)_{\rm sync}
+\left(\frac{\mathrm{d} \gamma}{\mathrm{d} t}\right)_{\rm comp}+
\left(\frac{\mathrm{d} \gamma}{\mathrm{d} t}\right)_{\rm ad}\right]
\end{equation}
The synchrotron radiation loss is the usual $\dot{\gamma}_{\rm sync}\sim \gamma^2 B^2$. The energy loss to inverse Compton radiation is calculated using the \citet{jones} kernel, which is a continuous approximation although electrons can lose significant amounts of approximation in discrete encounters with photons in the Klein-Nishina regime. Furthermore, the Jones kernel supposes an isotropic photon distribution. This should not be too bad an approximation as the angle-dependence of the cross-section is limited in the Klein-Nishina regime. 

The adiabatic losses for two dimensional expansion of relativistic material are
\begin{equation}
\left(\frac{\mathrm{d} \gamma}{\mathrm{d} t}\right)_{\rm ad}=\gamma \left(\frac{\mathrm{d} \ln n^{1/3}}{\mathrm{d} z}\right)=-\frac{2}{3}\frac{\gamma}{z}\frac{v}{R_s}
\end{equation}

The evolution of the physical parameters of the flow ($B$, $v$ etc) with $z$ is described in the next section. Once these are determined, the differential equation governing the evolution (Eq.~\ref{eqevol}) is numerically integrated along $z$ using a logarithmic grid in $\gamma$, following the scheme proposed by \citet{changcooper}. The results presented thereafter typically use at least 1000 points for the particle energy grid and 200 points in frequency $\nu$ to represent emission spectra. The results have been checked to be independent of the {numerical details}.

\subsection{Evolution of physical conditions in the flow\label{bernouilli}}

The evolution of the nebular flow speed $v$ and magnetic field $B$ with $z$ are computed by using the relativistic Bernouilli equation stating that
\begin{equation}
\frac{\Gamma}{n}\left(\frac{B^2}{4\pi \Gamma^2}+P+ne\right)= {\rm constant}
\end{equation}
along streamlines for stationary, adiabatic motion. $\Gamma$ is the bulk Lorentz factor of the flow $\Gamma=(1-v^2/c^2)^{-1/2}$. The additional relationships for the magnetic field $B$, density $n$, pressure $P$ and internal energy $e$ come from the conservation of (toroidal) magnetic flux $Bvz/\Gamma$, of the number of particles $nvz^2$ and the relativistic equation of state $P=n(e-mc^2)/3 \propto n^{4/3}$ \citep{kc1}.

Close to the shock, the total energy is dominated by pressure. The density is then roughly constant, the flow speed decreases as $v\propto 1/z$ and the magnetic field increases because of flux conservation {(Fig.~3)}. The flow becomes magnetically dominated at large radii ($z \ga 10$) at which point the density varies as $1/z^2$, the magnetic field decreases as $B\propto 1/z$ and the asymptotic flow speed is $\sigma c$ \citep{kc1}.

The present approach is admittedly simplistic. MHD simulations of a spherically symmetric pulsar wind interacting with a constant external medium exist \citep{bucciantini} and give insights into possible shortcomings. The shape of the pulsar termination shock is bullet-shaped rather than spherical. An asymmetric pulsar wind would surely change the aspect of the flow (as it does with the Crab and Vela pulsars). Shocked material flows along two channels, a fast one with material from the forward termination shock and a slower one with material from the backward facing termination shock. The nebular flow loses energy to radiation, contradicting the adiabatic approximation. Collimation shocks may re-energise particles along the jet; mixing can occur at the surface discontinuity between the fast-moving shocked pulsar wind and the slow-moving stellar wind. Mass loading will slow down if not disrupt the flow. At larger distances, the flow shears because of orbital motion (\S5.4), leading to additional complications. Finally, the flow merges with the stellar wind and/or ISM on scales of $\sim 0.01$~pc {based on the densities}.

\begin{figure*}[ht]
\resizebox{18cm}{!}{\includegraphics{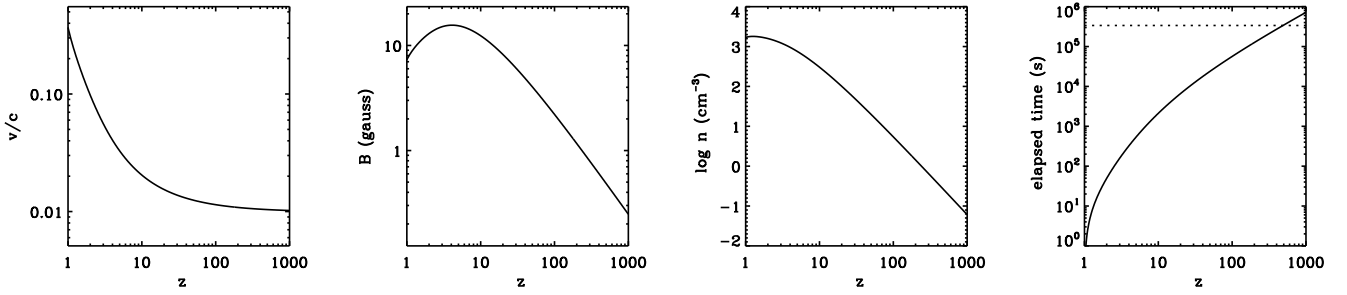}}
\caption{From left to right: evolution in the pulsar nebula of the flow speed, magnetic field and density as a function of the normalised distance $z$. The rightmost panel shows the time needed to reach a given distance $z$ in the flow. At $z\approx 800$ ($d=zR_s\approx 10$~AU), the elapsed time is equal to the orbital period (dotted line). The model parameters are $\dot{E}=10^{36}$~erg~s$^{-1}$, $\gamma=10^5$, $\sigma=0.01$ and $R_s=2\cdot 10^{11}$~cm, appropriate for \ls\ at periastron. \label{nebevol}}
\end{figure*}

\subsection{Calculated spectrum}
Neglecting any effect of the orbital motion (see \S\ref{appear}), the overall spectrum results from the summed contribution of the synchrotron and inverse Compton spectra (on stellar photons) calculated for subsections $dz$ of the nebula extending away from the star. As was done for energy losses, the particle distribution and radiation fields are assumed spatially homogeneous and isotropic {in each subsection}. The inverse Compton spectrum takes into account the reduction of the cross-section for interactions in the Klein-Nishina regime via the Jones kernel. The radiation field of the stellar companion is simply diluted by the inverse square of the distance to the star ($U_\star\propto d_\star^{-2}$ with $d_\star=d_s+(z-1)R_s$).

\begin{figure*}
\resizebox{!}{6cm}{\includegraphics{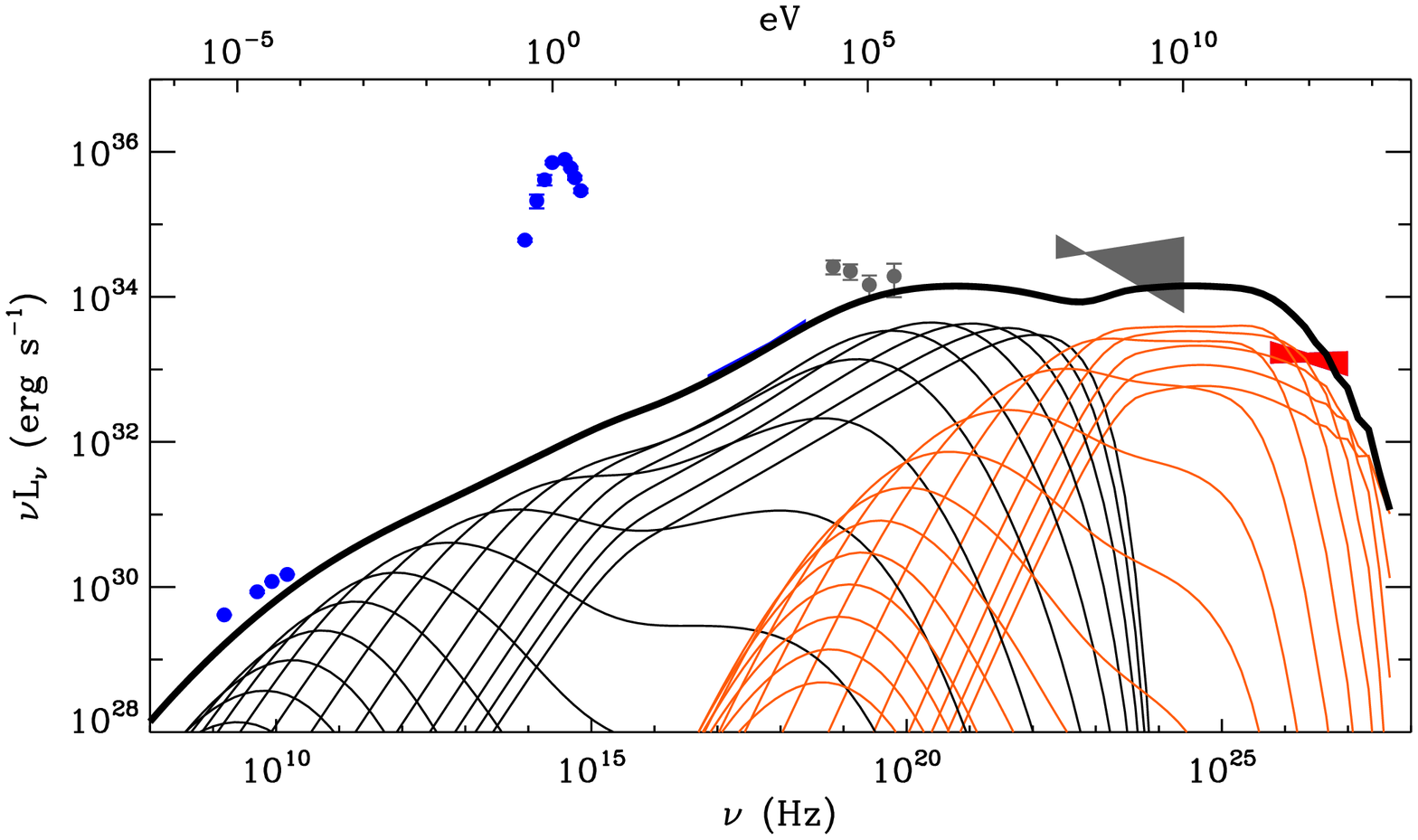}}
\resizebox{!}{6cm}{\includegraphics{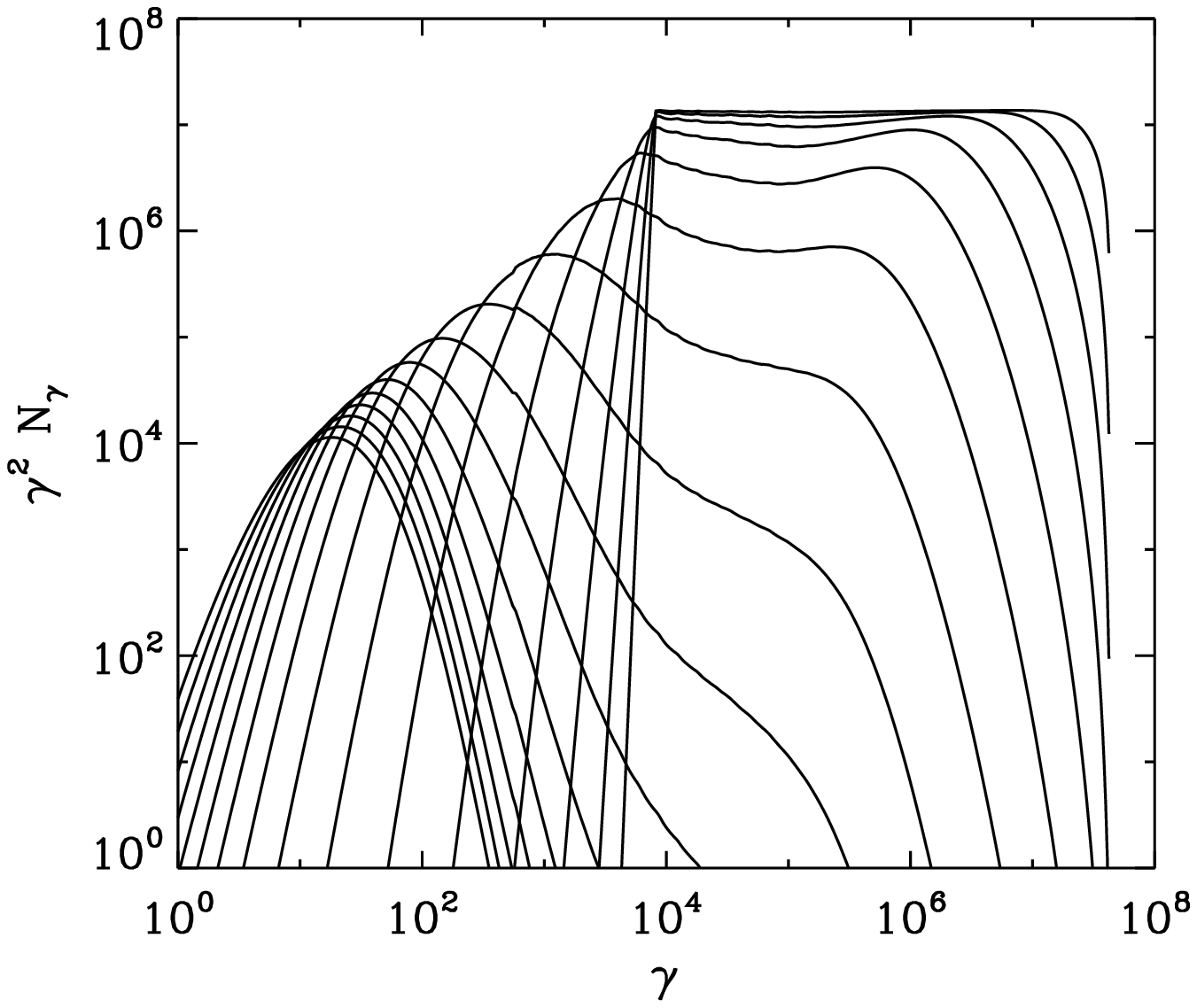}}
\caption{Left: Evolution of the emission along the pulsar nebula. The pulsar nebula is divided into sections at various distances $z$ with the corresponding synchrotron (black) and inverse Compton (grey) spectra shown. The first section shown is for $z=1$ ($dz=0.01$). The section $dz$ is then increased: the fifth spectrum is at $z\approx 1.5$, the tenth at $z\approx 15$, the last spectrum shown is at $z\approx1000$. The total spectrum of the pulsar nebula is shown as a thick black line. The model parameters are the same as in Fig.~\ref{nebevol}. The thermal emission at $\sim$1~eV is direct light from the stellar companion. Right: Evolution of the particle distribution as the shocked pulsar wind moves out from the system. The sections are the same as those shown in the SED. The initial $\gamma^{-2}$ power law cools rapidly at high energies due to synchrotron radiation. Adiabatic losses dominate at large distances. \label{evol}}
\end{figure*}

Fig.~\ref{evol} illustrates the evolution of the emission spectrum and particle distribution with increasing distance $z$ to the pulsar in the case of \ls. The first 3-4 spectra show the initial evolution towards the steady-state described in \S4. The particle distribution above $\gamma_S\sim 10^6$ cools rapidly at high energies due to synchrotron losses, resulting in a plateau down to $\nu_S$ as expected. Evolution then sets in around $z\sim 10$, with the spectral luminosity decreasing with $B$. The particle distribution steepens with emission sweeping the optical to IR bands. Further evolution, corresponding to peak emission in mm to radio, is set by adiabatic losses.

Self-Compton radiation is negligible. Synchrotron emission was optically thin down to low radio frequencies in all the models computed for \S6. This is a consequence of the flow geometries, which are discussed in \S\ref{appear}. Still, synchrotron self-absorption might be possible in a flow directed predominantly along the line-of-sight.

In the absence of orbital motion, the spectrum seen by the observer is simply the integrated emission along the `comet' tail nebula, just as for isolated pulsar interacting with the ISM. However, as can be deduced from Fig.~\ref{evol}, the radio emission originates from regions several AU away from the pulsar ($z\ga 100$). The time $\Delta t$ it takes for particles to reach this location, about a day here, is a significant fraction of $P_{\rm orb}=3.9$~days. Orbital motion will have changed the inner boundary condition so the spectrum seen by an observer will be a composite of SEDs from all orbital phases $\phi$. Here, the only parameter dependent on the orbital phase is the pulsar standoff point $R_s$. Other dependences can be envisioned if, for instance, the pulsar wind is anisotropic or if the magnetization parameter is a function of distance to the light cylinder (with higher $\sigma$ expected close to the pulsar).

The relevant timescales to compare are the flow timescale $\tau_{\rm flow}=d/v$ to reach a distance $d$ {from the shock} and the orbital motion timescale $\tau_{\rm orb}=d_s/v_{\rm orb}$. The orbital timescale can vary greatly for a highly eccentric orbit. Nevertheless, taking $v$ equal to its asymptotic value $\sigma c$, the distance where orbital motion becomes important will on average be
\begin{equation}
d_{\rm n}\approx \sigma c P_{\rm orb} / 2\pi
\label{dorb}
\end{equation}
and the associated timescale, $P_{\rm orb}/ 2\pi$. As it turns out, the synchrotron emission from \ls\ depends little on the orbital phase because $R_s/d_s$ is constant with $\phi$. The spectrum {seen by an observer should} therefore be close to the integrated spectrum shown in Fig.~\ref{evol} even at low frequencies (see \S\ref{ls5039}).

\subsection{Appearance of the cometary nebula\label{appear}}
On a small scale, the shocked material flows away from the binary on a straight path, following the direction given by the vector difference of the stellar wind and orbital speeds $\mathbf{v}_{\rm w}-\mathbf{v}_{\rm orb}$. On a larger scale, the material `bends' as the orbital motion becomes important at $\sim d_{\rm n}$ {(about 1 AU for \ls;} Eq.~\ref{dorb}). Parcels of material are still on outward straight paths but the orbital motion shears the nebula. The opening angle of the flow ($\sim R_s/d_s$) is small so that the width of the `comet' tail is smaller than $d_{\rm n}$. For a circular orbit, the large scale appearance is then an Archimedes spiral with a step size $2\pi d_{\rm n}=\sigma c P_{\rm orb}$.

More generally, the flow speed amplitude and direction change along the orbit and the exact shape can differ from a perfect spiral. Furthermore, emission along the spiral is not constant but depends upon the distance $d$ to the pulsar or, alternatively, upon the time elapsed since the particles left the pulsar. The flux received by the observer may also be Doppler by factors of a few (de)boosted initially, when the flow speeds are high (neglected here). Finally, the orbital orientation will also change the appearance of the nebula. For an edge-on system ($i=90\degr$), the nebula will mimic a microquasar bipolar jet.

The trajectory of material leaving the pulsar wind termination shock can be found as a function of the elapsed time $t$, using polar coordinates $(r,\theta)$ in the orbital plane. The distance $d$ reached along the nebula is
\begin{equation}
d=\int_{0}^{t} v dt
\end{equation}
where $v$ is the flow speed calculated in \S\ref{bernouilli}. The radius $r$ with the coordinates centered on the massive star is given by
\begin{equation}
r^2=d^2+2 d d_s \cos \beta+d_s^2
\end{equation}
where $\beta$ is the angle at which the shocked material flows and $d_s$ is the star-pulsar separation at $t=0$. If $\mathbf{e_v}$ is the unit vector along the flow direction and $\mathbf{e_r}$ is the radial unit vector from the massive star to the pulsar location at $t=0$, then  $\cos \beta= \mathbf{e_r} . \mathbf{e_v}$. The angle $\theta$ of the material is then
\begin{equation}
\theta=\omega+\eta-\psi
\end{equation}
where $\omega$ is the periastron argument, $\eta$ is the true anomaly of the pulsar at $t=0$ (angle measured from the star center, $\eta=0$ at periastron) and $\psi$ is derived from $r\sin \psi= d \sin \beta$. 

The full pattern follows from having $\eta$ vary with time $t$ according to the usual relationships governing orbital motion ({\em i.e.} using the eccentric anomaly). Note that $v$, $d_s$ and $\beta$ also vary with orbital phase. The result is then projected onto the plane of the sky, given the orbital inclination $i$.

\begin{figure}
{\includegraphics[width=\columnwidth]{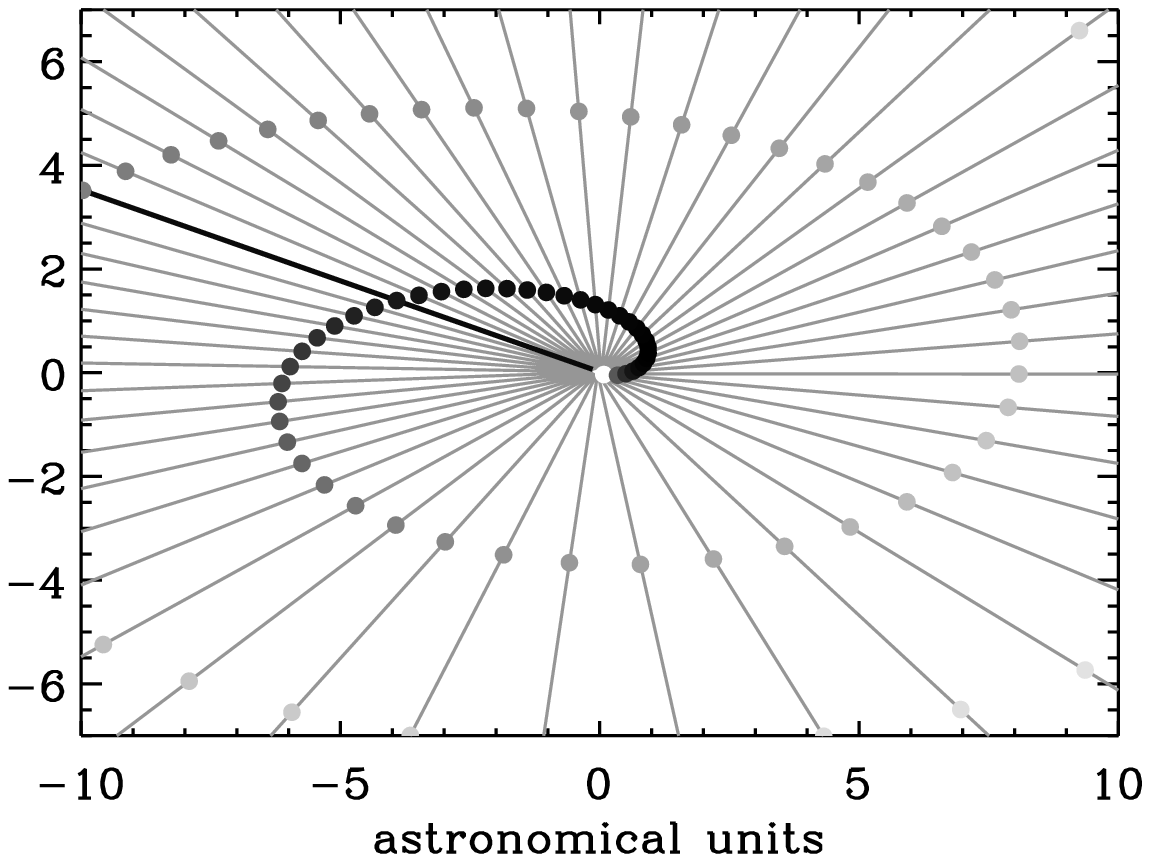}}
\caption{Large scale appearance of the pulsar nebula around \ls\ at periastron. In practice, limited angular resolution smoothes out the emission from what is shown here (see Fig.~\ref{ls5radio}). The binary is located at the origin. Its size of a few tenths of AU is much smaller than the scale of the map. Material originating from the standoff point travels outwards following the direction of the comet tail at each orbital phase (shown by grey lines; {the projected outflow path at apastron is indicated by a dark line}). Because of the orbital motion, the emission winds up into a spiral. To illustrate this, dots are plotted at the locations reached by material expelled at {\em regular} time intervals along the orbit. The total elapsed timescale represented corresponds to almost an orbital period and a half. The greyscale codes the intensity of the 5~GHz radio emission ({square root} scale). At large distances, the step of the spiral tends to $\sigma c P_{\rm orb}$ which is $\approx 7$~AU for the adopted parameters (same as those of Figs.~\ref{nebevol}-\ref{evol}). {The inclination ($i=120\degr$, corresponding to a retrograde pulsar orbit, {see \S6.1.2}), and the major axis orientation ($\omega=226\degr$) used here are consistent with the measurements of the companion star's radial velocity} (see orbit schematic in Fig.~6 of \citealt{casares}). \label{spiral}}
\end{figure}

Fig.~\ref{spiral} shows the spiral pattern computed for \ls\ with $\dot{E}=10^{36}$~erg~s$^{-1}$, $\gamma=10^5$, $\sigma=0.01$ and the pulsar located at periastron. The figure represents the emission from particles `ejected' over a timescale of about an orbital period and a half. The spiral `turns' around and outwards as the pulsar moves along the orbit. The lines illustrate the direction at which material flows out from the binary. Spacing between lines correspond to equal intervals of time; in other words, lines are more spaced out at periastron due to fast orbital motion {(before projection on the plane of the sky)}. For very eccentric orbits (such as \psrb, see \S\ref{psrb}), material will preferentially flow out in the direction of the major axis on the apastron side (for a radial outflow). There will be a preferred emission axis, even when the system is seen pole-on. Another effect illustrated by Fig.~\ref{spiral} is the bunching of lines due to the projection on the sky plane. A binary with a circular orbit but seen at some inclination will also have a preferred emission axis.

Besides the orbital eccentricity and inclination, a third effect that will shape the appearance of the cometary nebula is the variation of emission with orbital phase. The amplitude can change with orbital phase. But even if the spectrum varies little along the orbit, the locus of peak emission changes as the elapsed time depends directly on the absolute scale $R_s$ (hence phase). If the standoff point is twice as far away at apastron than it is at periastron, then the spatial scale on which the radio peaks also doubles. In Fig.~\ref{spiral}, the intensity of the 5~GHz radio emission is represented by the grayscale coloring. Note how far out material on the {upper} left hand side (corresponding to apastron) has a higher intensity than material on the right hand side (periastron) although the latter was emitted a shorter time ago. 

All these effects combine to create radio maps that are far from trivial (see \S6). Despite the complexity, there is one simple prediction: the step of the spiral should be $\approx \sigma c P_{\rm orb}$. {In principle}, the radio nebula of $\gamma$-ray pulsar binaries could be used to measure the average magnetization parameter of relativistic pulsar winds. The particular cases of \ls, \lsi\ and \psrb\ are examined in the next section.

\section{Application to the observed $\gamma$-ray binaries}

\subsection{\ls \label{ls5039}}
The wind of the O6.5V companion star in \ls\ is assumed to be radiatively-driven with a velocity and density given by
\begin{equation}
v_w=v_{\infty} \left(1-\frac{R_\star}{d_s}\right)^{\beta} \label{beta}\\
\rho_w=\frac{\dot{M}_w}{4\pi d_s^2 v_w}
\end{equation}
with $\beta\approx1$. Spectroscopic observations of \ls\ constrain $\dot{M}_w$ to $\approx 10^{-7} M_\odot$~yr$^{-1}$ and $v_\infty\approx 2400$~km~s$^{-1}$ \citep{mcswain04}. The wind is assumed isotropic and purely radial. Since the orbit of the pulsar takes it to almost a stellar radius of the companion at periastron in \ls\ and \lsi, more complex wind launching conditions  than those assumed above could play a role in setting $R_s$. A significant wind rotation might also influence the location of the standoff point {and the flow direction}.

The pulsar spindown power was taken to be $\dot{E}=10^{36}$~erg~s$^{-1}$, close to that measured in \psrb\ (\S\ref{psrb}). $R_s$ then varies between 2~10$^{11}$~cm (periastron) and 4~10$^{11}$~cm (apastron). The two parameters left are the pulsar wind magnetization $\sigma$ and Lorentz factor $\gamma_p$. Low values of $\sigma\ga 10^{-4}$ yield low downstream magnetic fields. Hence, the inverse Compton emission dominates with far too little X-rays emitted to match the SED. At the other end, values of $\sigma\approx 0.1-1$ lead to overall emission inefficiency because electrons move quickly out into regions of low radiation fields (the downstream flow speed is high). Values of $\sigma$ in the 0.001-0.01 range yield the best results. However, with $\sigma\approx 0.001$, the characteristic size of the radio emission (Eq.~\ref{dorb}) would be too small by a factor 10 to match the observed scale of $\sim 10$~AU \citep[][see below]{paredes}. The best compromise was $\sigma=0.01$. Coincidentally, the downstream flow speed saturates at a value ($\sim 3000$ km~s$^{-1}$) that is then close to the terminal velocity of the stellar wind ($\sim 2000$ km~s$^{-1}$).

The Lorentz factor $\gamma_p$ influences the lower cutoff of the particle distribution. Lower values enable the inverse Compton emission to extend down to the GeV range, as discussed in \S\ref{simplesed}. With more low energy electrons, the radio emission is also more easily accounted for. A value of $\gamma_p=10^5$ was the best compromise: at $10^6$ the GeV emission is unaccounted for, at $10^4$ the radio emission is overpredicted and the downstream particle distribution then extends more than four decades in energy. The adopted values are $\sigma=0.01$ and $\gamma_p=10^5$.

\subsubsection{The SED}
The evolution of the SED as the material flows away from the pulsar is shown in Fig.~\ref{evol}. The high energy SED is very much that described in \S4, with the spectral breaks in the synchrotron and inverse Compton spectra at their expected locations. Electrons in the comet tail enable the summed emission to extend down to low frequencies and also contribute to emission in the GeV domain. {The comparison to the observations shows reasonable but not perfect agreement}. Changes in the particle distribution, stellar or pulsar wind could reduce the discrepancies but have not been attempted here. The purpose {at this stage} is to validate the model without fine-tuning and to compare the $\gamma$-ray binaries under similar assumptions.

\begin{figure}
{\includegraphics[width=\columnwidth]{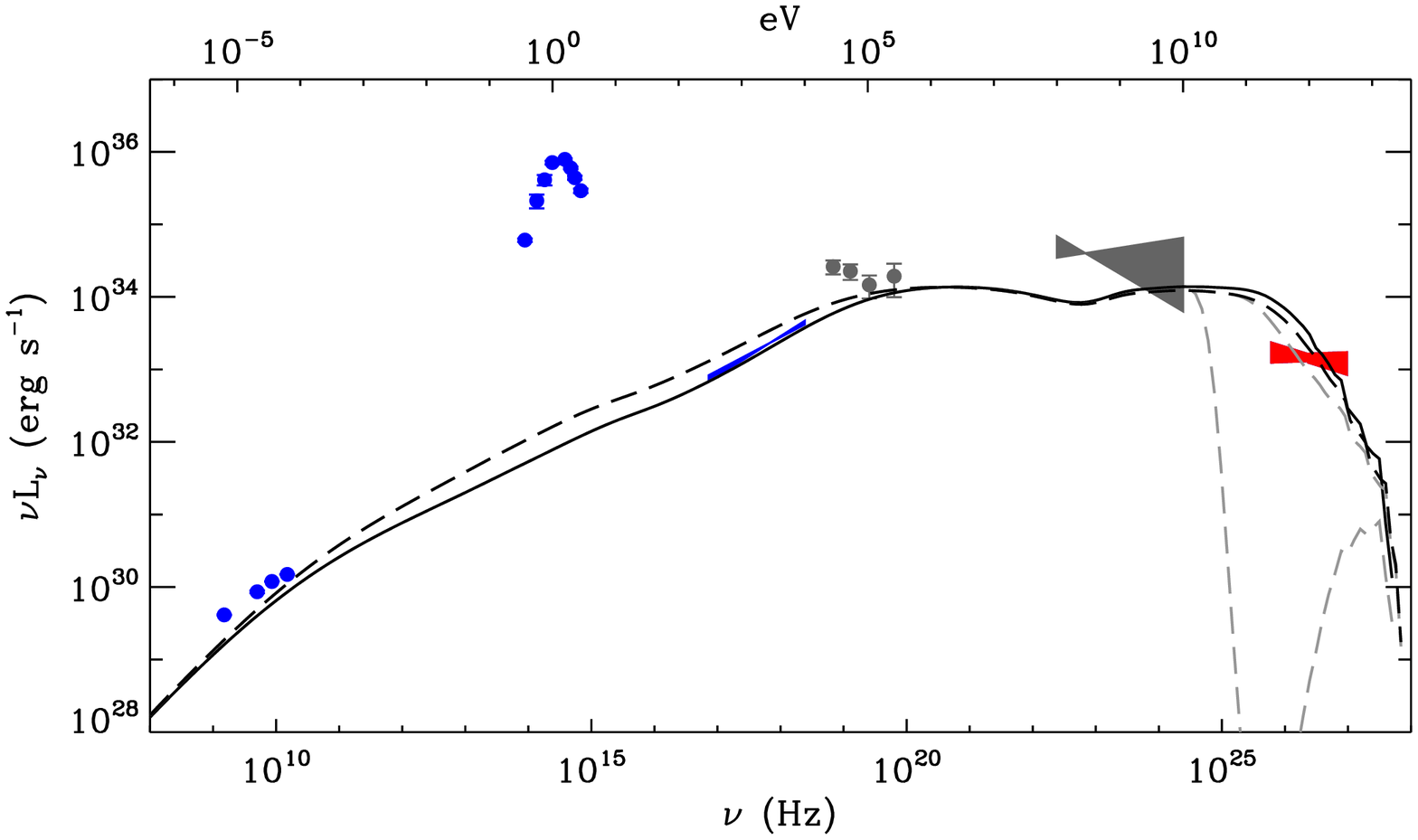}}
\caption{Spectral energy distribution of \ls\ at periastron (dark full line) and apastron (dark dashed line).  {The corresponding grey lines show the fraction received by the observer after  absorption of the VHE $\gamma$-rays on stellar photons}. Parameters are identical to those of Figs.~\ref{nebevol}-\ref{evol} (but using $R_s=4\cdot10^{11}$~cm at apastron). {Changes in the overall SED are limited and only minor variability is expected except at VHE energies where $\gamma\gamma$ absorption strongly modulates the flux on the orbital period. }\label{orbsedls5}}
\end{figure}

Fig.~\ref{orbsedls5} shows the variation of the SED between periastron and apastron. Apart from the slight lowering of the synchrotron break frequency (see \S\ref{simplesed}), the spectral energy distributions are nearly identical. This agrees with the lack of strong orbital variability observed in X-rays or radio. {However, an orbital modulation is unavoidable at TeV energies because of $\gamma\gamma$ absorption (see below).}

At $\gamma$-ray energies, the model has plateau emission across the {EGRET} band and up to the \hess\ regime, as expected in \S4. However, (1) the GeV flux is underestimated and (2) the high energy break is a factor 2-3 too low than what would be required to match the hard \hess\ power-law ($\sim 2$ TeV). \ls\ is in the Galactic Plane so that the {EGRET} flux could be overestimated {(i.e. should be interpreted as an upper limit)} if the diffuse $\gamma$-ray background subtracted is underestimated or if several unresolved sources contribute to the emission. That the measurements are non-contemporaneous should not be a problem in the present scenario, where variability is expected to be small (except {for the possible pair cascade}, see below). 

The approximate treatment of the emission geometry may be responsible for these discrepancies.  Adjusting $\dot{E}$, $\sigma$ or the standoff point distance can alleviate both problems (Eq.~\ref{comps}). Tweaking the accelerated distribution of particles might help, but lacks justification. There could be a direct contribution from acceleration processes in the pulsar magnetosphere. Two other avenues can also be explored.

First, the anisotropy of the stellar radiation field might play a role. The outgoing energy of an up-scattered photon depends on the interaction angle with the higher energies reached at $\pi$. Even if the distribution of electrons is isotropic, the mean angle of interaction of the photons detected by the observer will depend on the orbital phase, hence there should be an energy effect. This is likely to be small at TeV energies as there is not much energy dependence on angle in the Klein-Nishina regime \citep{moderski}. 

A second important effect is $\gamma\gamma$ absorption \citep{dermer,dubus,bednarek}. High energy $\gamma$-rays are emitted in the immediate vicinity of the pulsar where they can then interact with stellar photons to produce $e^+e^-$ pairs. The threshold $\gamma$-ray energy is $\approx$ 30~GeV. {The transmitted flux is shown in Fig.~\ref{orbsedls5} as grey lines for both periastron and apastron. The calculation assumes that $\gamma$-ray emission is isotropic and occurs at the pulsar location (see \citealt{dubus} for details). The opacity depends on orbital phase, with a maximum close to periastron passage when most of the $\gamma$-ray emission between 0.1-10~TeV should be absorbed. An orbital modulation of the VHE flux measured by HESS is therefore expected. Note that this is not a discriminating feature with the accretion/jet scenario since any $\gamma$-ray emission occurring within $\approx 1$ AU of the companion star in this system (for instance, in regions close to the black hole) would be significantly modulated. }

The average result of absorption will be a hardening of the high energy {(HESS range)} spectrum. Pairs created by the absorption of $\gamma$-rays can subsequently upscatter stellar photons to very high energies, thereby initiating a cascade {(not taken into account in the calculation shown in Fig.~\ref{orbsedls5})}. This can significantly increase the emission below 30~GeV as the absorbed energy is redistributed below the interaction threshold {if the cascade is very strong} \citep{aharonianls}. This would also introduce orbital variability below 30~GeV \citep{dubus,bednarek}. {However, note that high energy pairs above $\gamma_S$ will radiate predominantly synchrotron emission instead of upscattering photons (\S\ref{timescales}), reducing the extent of the cascade} \citep{aharonianls}.

Understanding the exact impact of these two effects will require complex modelling, including a better representation of the shock geometry. The conclusion is that the present fairly rustic model is already able to account for the whole SED adequately, in the manner described in \S4 using straightforward arguments.

\subsubsection{Radio maps}
The total radio intensity does not vary much along the orbit but the time/distance at which maximum radio emission is reached does depend on the orbital phase. The evolution of the radio flux with distance to the pulsar was calculated as above for various orbital phases and used together with the geometric model for the nebula (\S\ref{appear}) to create emission maps of the system. Free-free absorption on the stellar wind becomes optically thin at $\sim 1$ AU (see rough estimate in \S\ref{pulsar}), close to the distance at which radio emission starts. Hence, this effect is neglected.

\begin{figure*}
\resizebox{4.4cm}{!}{\includegraphics{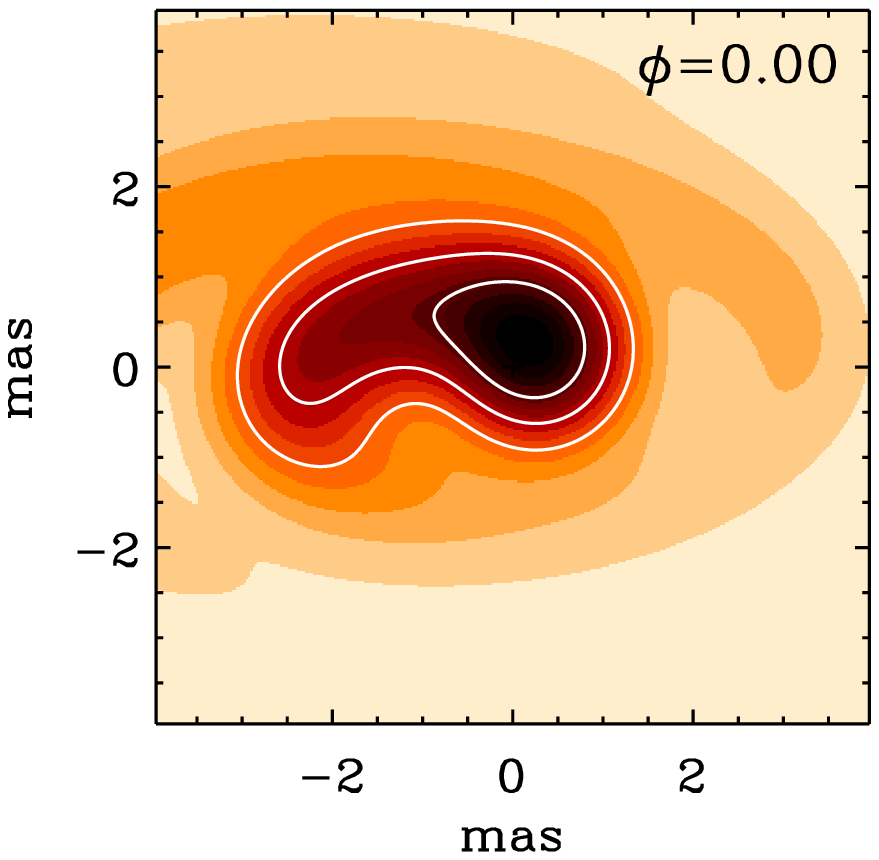}}
\resizebox{4.4cm}{!}{\includegraphics{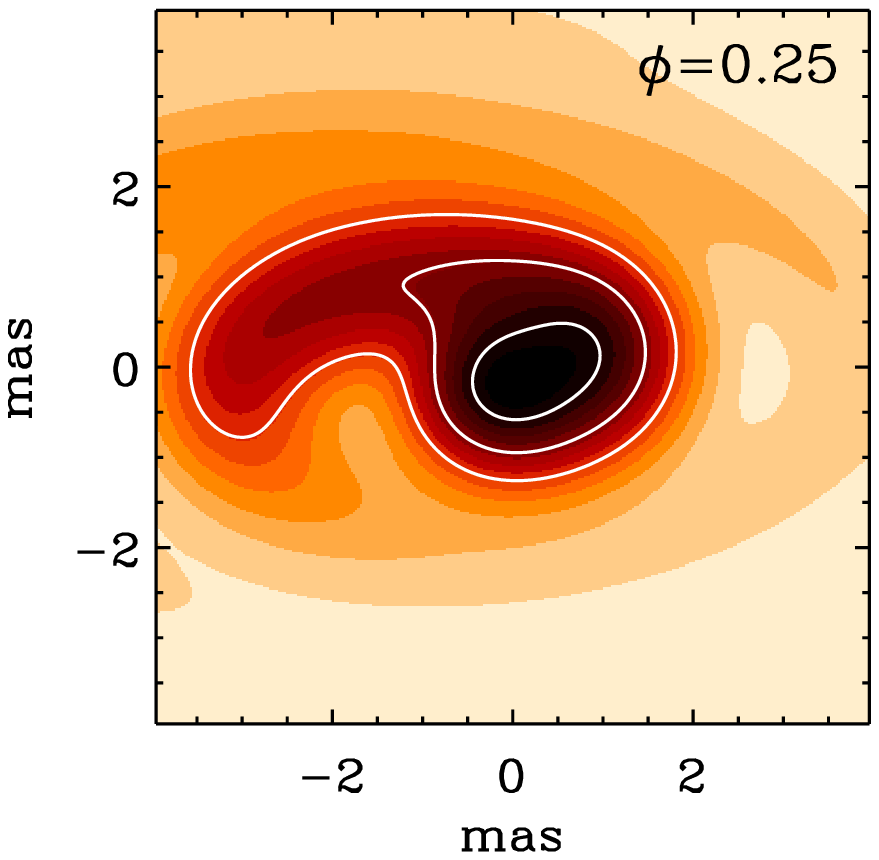}}
\resizebox{4.4cm}{!}{\includegraphics{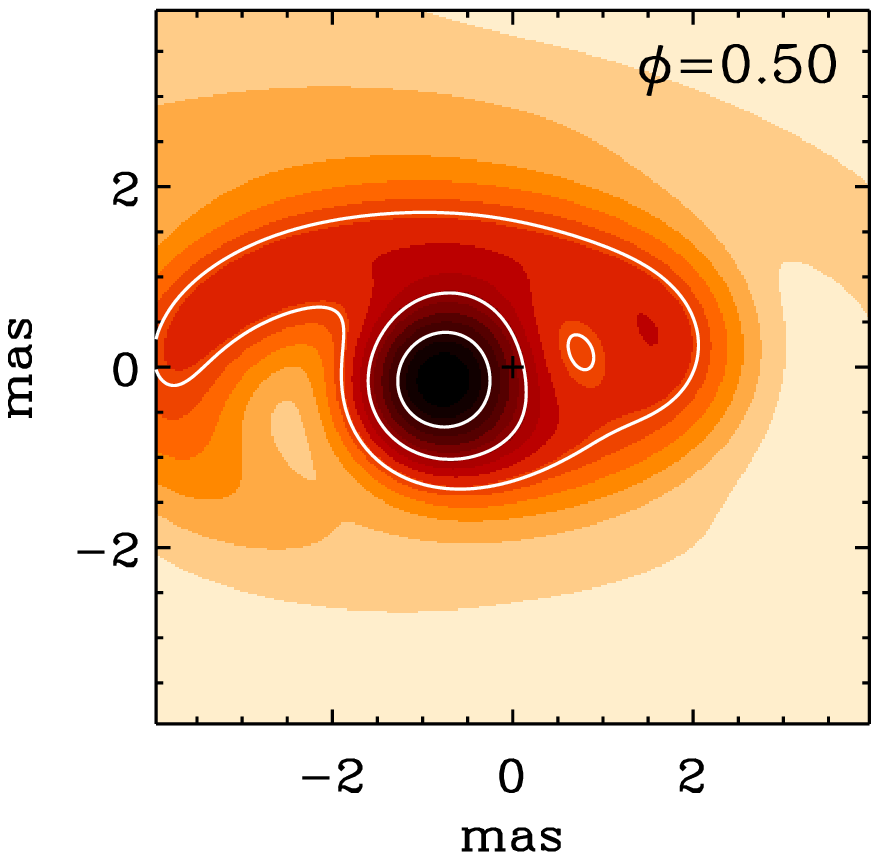}}
\resizebox{4.4cm}{!}{\includegraphics{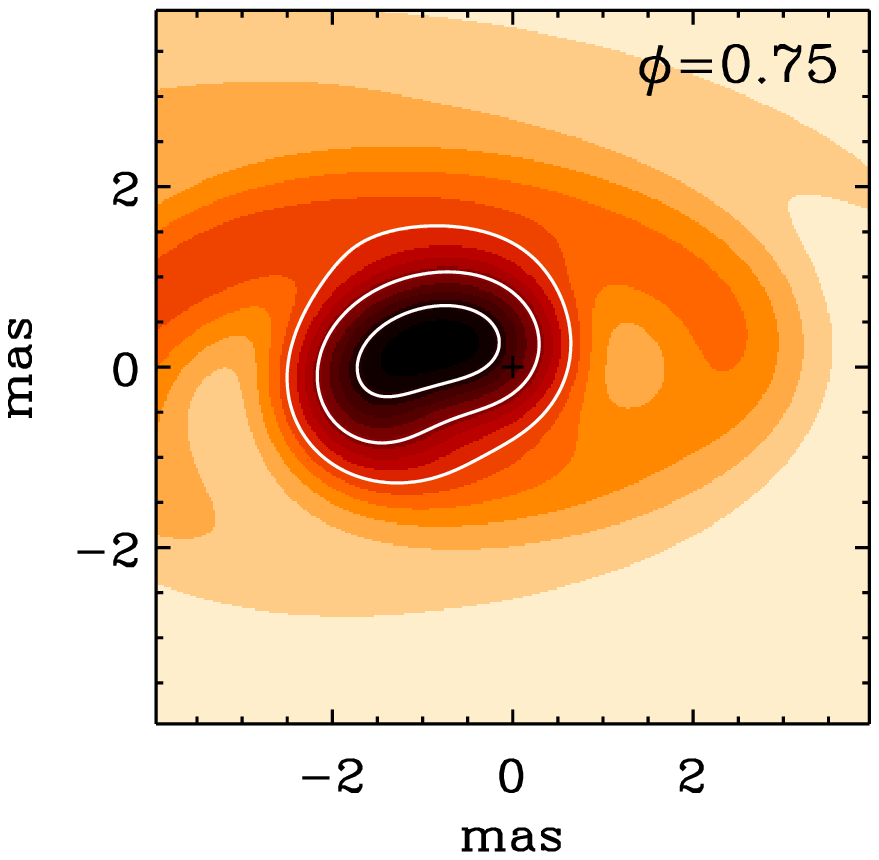}}
\caption{Orbital evolution of the 5~GHz radio emission from the pulsar nebula in \ls. The maps were composed by summing up the contributions along the spiral nebula, as calculated for the various orbital phases $\phi$ (see Fig.~\ref{spiral} for the $\phi$=0 case). The maps were then smoothed by a gaussian kernel with a Full-Width at Half Maximum (FWHM) of 0.8 mas (3 $10^{13}$ cm at 2.5~kpc). The maps share a common color scale to allow direct comparison, with contours at 50\%, 25\% and 10\% of the maximum orbital radio emission. The maps at $\phi$=0-0.25 are {very} similar to that observed by \citet{paredes}. \label{ls5radio}}
\end{figure*}

Fig.~\ref{ls5radio} shows the evolution of the 5~GHz emission as a function of orbital phase $\phi$. The first map (periastron $\phi=0$) shown uses the same parameters as those in Fig.~\ref{spiral}, except that the emission is smoothed by a Gaussian with FWHM of 0.8~mas to approximate the observing setup of \citet{paredes}. The system is seen at an inclination $i=120$\degr, i.e. 30$\degr$ below the orbital plane ($i=60$\degr\ would be 30\degr\ above the plane, the degeneracy from the radial velocity is lifted by the comparison of {the curvature in the map to the VLBI observations}). The contours at 50\%, 25\% and 10\% of the maximum surface intensity can be compared to the 5~GHz VLBI observation. The maps at $\phi=0-0.25$ are strikingly similar to that observed by \citet{paredes}, with slightly curved emission extending a few mas and the inner emission pointing in a slightly different direction. The observed very weak counter-jet would be associated with emission emitted at the previous periastron passage.

The model predicts changes in the radio maps with $\phi$ (Fig.~\ref{ls5radio}) that can be verified by further VLBI observations. An animated map gives the distinct impression that blobs are alternatively ejected to the sides of the system, cunningly mimicking a microquasar. Accurate astrometry should show that the position of peak radio emission varies by a few mas. The orbital motion produces a `comma' shape that is more pronounced at $\phi\approx 0$, when particles emitted at {\em apastron} start emitting in the radio band. Emission occurs principally along the major axis of the sky-projected system orbit. This effect shows up as a bunching of the lines representing the trajectories in the mid-plane of Fig.~\ref{spiral}.

On progressively larger scales the emission becomes egg-shaped along the same axis, with the tip of the `egg' on the apastron side as projected onto the sky. MERLIN and EVN radio observations do resolve emission on a larger scale on one side than on the other \citep{paredes02}. The observed emission is more linear than in the model with $i=60\degr$. A greater inclination would help since emission becomes exactly line-shaped for an edge-on orbit $i=90\degr$. {However, the X-rays would then be eclipsed when the pulsar moves behind the companion; an upper limit using the orbit and assuming the X-ray emitting region is a sphere of radius $R_s$ gives $\approx 75\degr$}. The observed extent of radio emission is also larger than expected. On an equivalent map to the EVN observation, the  5~GHz intensity decreases to 1\% of the peak value at 12~mas on one side and 6~mas on the other instead of 35~mas and 25~mas. The quantitative agreement is not as good as on scales $\sim \sigma c P_{\rm orb}$. 

While the map in Fig.~\ref{ls5radio} shows basically only emission within an orbital timescale, maps on scales  $\gg \sigma c P_{\rm orb}$ sum up emission from dozens of orbits, representing only a tiny fraction of the initial energy budget. The physics on such scales is undoubtedly more complex than the simple approach taken in \S\ref{bernouilli}: mixing occurs with the surrounding medium; recollimation shocks re-energise the electrons; shearing introduces turbulence; magnetic field changes as it is azimuthally stretched; shock conditions at earlier times were different etc. Note that the outflow length of $\sim$500~AU in \citet{paredes02} coincides roughly with the distance at which the stellar wind merges with the ISM. Incidentally, many of these complications are also encountered in the propagation of relativistic jets. 

The conclusion is that the pulsar model provides a very good quantitative explanation for the radio emission emitted on scales of AU in \ls\ and provides a qualitative framework to understand radio emission on larger scales. Radio observations have tremendous potential to validate the model.

\subsection{\lsi}

The pulsar $\dot{E}$ and $\gamma_p$ were taken similar to those of \ls\ for comparison purposes. A slightly higher value of $\sigma=0.02$ was adopted here {to better match the observed radio scales (\S6.2.4; note that Table 2 uses $\sigma=0.01$)}. 

\subsubsection{Stellar Wind\label{wind}}
The stellar companion in \lsi\ is a B0Ve type star which significantly complicates modelling over the case of \ls. Ultraviolet and infrared observations of Be stars show that their stellar wind has two components: a fast, polar outflow with properties typical of radiatively-driven winds; and a slow, equatorial outflow \citep{waters1986}. 

A typical polar wind was modelled using Eq.~\ref{beta} with $\dot{M}_w= 10^{-8}~M_\odot$~yr$^{-1}$ and $v_\infty= 2000$~km~s$^{-1}$ \citep{waters1988}. The standoff point $R_s$ then varies between 8~10$^{11}$~cm and 4~10$^{12}$~cm. The equatorial wind was modelled as
\begin{equation}
v_w=v_0 (R/R_\star)^{n-2}\\
\rho_w=\rho_0 (R/R_\star)^{-n}
\end{equation}
with $n\approx3$, $v_0\approx 10$~km~s$^{-1}$ and $\rho_0\approx 10^{-11}$~g~cm$^{-3}$ from the comparison to infrared observations of \lsi\ \citep{waters1988,Marti95} . This corresponds to an equatorial mass flux about a 100 times larger than the polar mass flux. The typical opening angle of the Be star equatorial outflow is 15\degr. H$\alpha$ observations show the disc extends to 5-20 $R_\star$ and that the terminal velocity is a few hundred km~s$^{-1}$. The standoff point is then between 5~10$^{10}$~cm and 2~10$^{12}$~cm, smaller than the polar wind values because of the higher densities. The major difference between the two types of stellar winds will be at periastron.

\subsubsection{The SED}
\begin{figure}
{\includegraphics[width=\columnwidth]{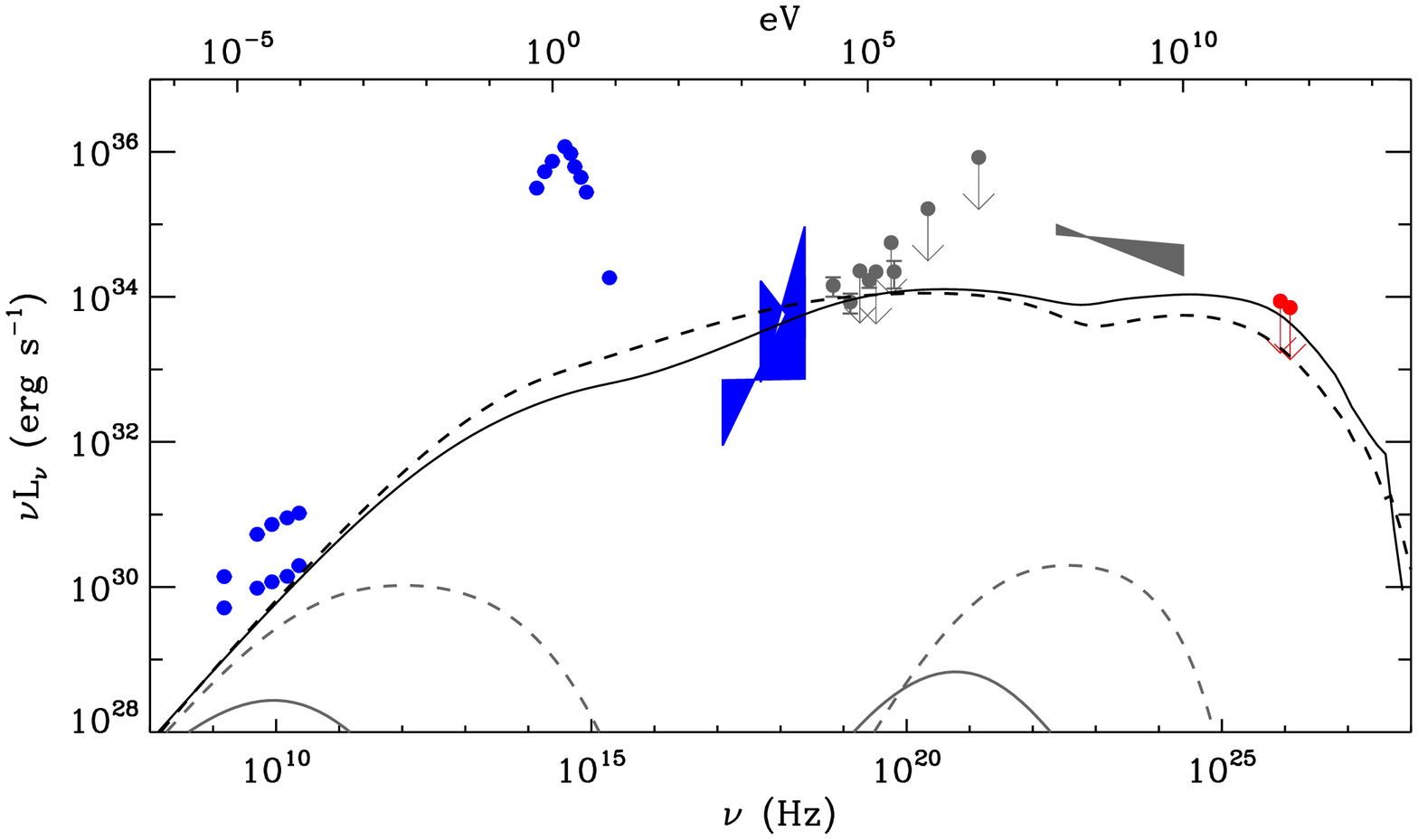}}
{\includegraphics[width=\columnwidth]{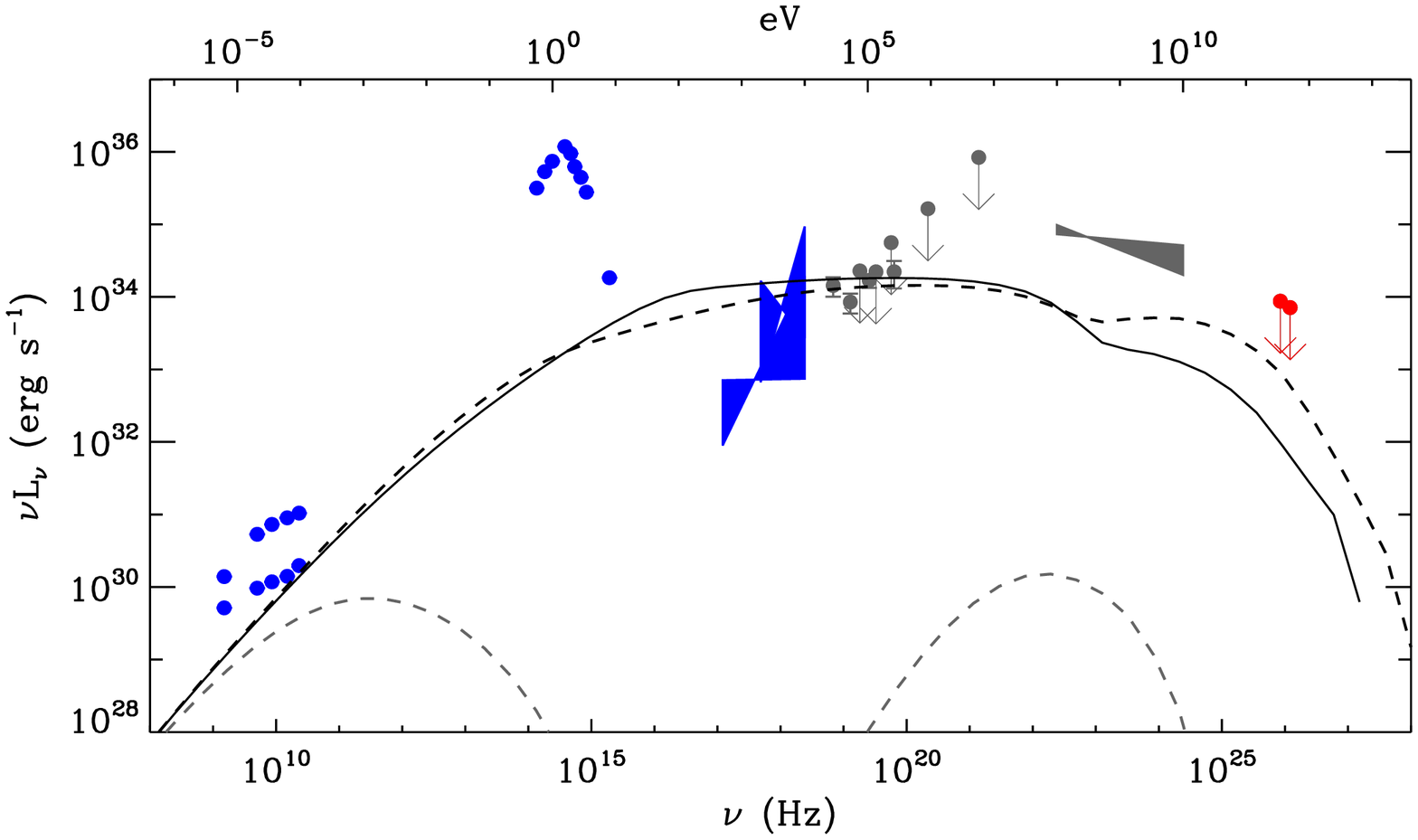}}
\caption{Spectral energy distribution of \lsi\ at periastron (full black lines) and apastron (dashed black lines). The pulsar is assumed to have $\dot{E}=10^{36}$~erg~s$^{-1}$ and $\gamma=10^5$ as for \ls, but a slightly higher $\sigma=0.02$. Top: The stellar wind is polar with a standoff point at $R_s=8\cdot 10^{11}$~cm (periastron) and $4\cdot10^{12}$~cm (apastron). Bottom: The stellar wind is equatorial with $R_s=5\cdot 10^{10}$~cm (periastron) and $2\cdot10^{12}$~cm (apastron). Grey lines show the fraction of the emission occurring when $\Delta t>P_{\rm orb}$ {(the periastron line is visible in the top panel but not in the bottom panel)}.\label{lsimodsed}}
\end{figure}

The resulting SEDs at periastron and apastron for both stellar winds are shown in Fig.~\ref{lsimodsed}. They agree well with the expectations set out in Table~\ref{freq}, most notably the much lower break frequencies for the equatorial outflow compared to the polar wind due to the smaller $R_s$.

Orbital variations in X-rays are expected to be greater than in \ls. Indeed, observations show variations of a factor $\sim$10 in soft X-rays (0.1-2~keV) and of smaller amplitude at higher energies. The maximum is inferred to be close to periastron \citep{harrison,casares2}. Consulting Fig.~\ref{lsimodsed}, this would seem to favor combining a small $R_s$ at periastron (equatorial wind) and a large $R_s$ at apastron (polar wind). At apastron, the pulsar is 15~$R_\star$ away from its stellar companion, in a region where the Be equatorial disc could already be truncated. 

At $\gamma$-ray energies, the TeV fluxes are consistent with the {Whipple} upper limits. Small $R_s$ for the equatorial flow result in high magnetic fields, conversely lowering the $\gamma$-ray emission. Non-intuitively, TeV emission will be higher at {\em apastron} than at periastron if the pulsar plows through a dense equatorial outflow close to its companion and a more tenuous wind further out. With $R_s$ closer to the pulsar, synchrotron losses largely dominate at periastron over IC emission. This can be tested by observations with {VERITAS} {or {MAGIC}}. As with \ls, the GeV fluxes are too low to match the {EGRET} spectrum. An $e^+e^-$ cascade is unlikely to play a role here as the $\gamma\gamma$ opacity is small except very close to periastron \citep{dubus}. 

At low energies, the radio emission is too steep, primarily because of the value of $\sigma$. A lower value ($\sigma\approx 0.001$) would yield better agreement: the downstream magnetic field is then lowered, reducing synchrotron losses and enabling stronger radio emission at large $z$. {However, the physical extent of the associated radio emission would be smaller. A compromise would be to have $\sigma$ change with orbital phase. The distance of the shock to pulsar $R_s$ changes by an order-of-magnitude here (compared to a factor 2 in \ls) and $\sigma$ could be envisioned to decrease as $R_s$ increases.}

\subsubsection{Radio outbursts}
\lsi\ has regular radio outbursts during which the flux increases by an order of magnitude. The outbursts last several days, peaks at an orbital phase $\approx 0.2-0.7$ and repeat on the orbital period. There is also an X-ray orbital modulation (a factor 10 in the {\em ROSAT} band, less above)  that precedes the radio peak by $\Delta\phi\approx 0.1-0.5$ (the exact delay varies between outbursts). \psrb\ has similar outbursts at periastron, associated with the pulsar crossing the Be equatorial outflow \citep{melatos}. Indeed, the Be nature of the companion is the main feature distinguishing \psrb\ and \lsi\ from \ls, which has no radio outbursts. 

Assume the pulsar samples the dense equatorial disc around periastron, and a more tenuous, fast wind at apastron. Changes in $R_s$, possibly $\sigma$, are expected along the orbit; yet these do not suffice to explain the outbursts because of the timescales involved. If such was the case, the $\sim 10$-day delay in the phasing of the X-ray and radio modulations would presumably be associated to the time it takes for electrons to cool from injection. The strongest radio emission would occur at apastron, when the shock is far from the pulsar and $\sigma$ is expected to be lower. The elapsed time to radio peak is then greater than the orbital period (grey line in Fig.~\ref{lsimodsed}) and this would smooth out the flux modulation. 

The impact of the Be star equatorial outflow is probably somewhere else. With a slow wind, the standoff point $R_s\approx 5\cdot 10^{10}$~cm is very small at periastron. Moreover, the equatorial wind is very slow so that the Bondi accretion radius $\propto 1/v_w^3$ (\S\ref{pulsar}) is actually greater than $R_s$. For the adopted parameters, the accretion radius is $\approx 2\cdot 10^{11}$~cm at periastron. Matter falls towards the compact object at a rate $\approx 10^{17}$~g~s$^{-1}$, but is not accreted: the pulsar wind is still able to hold off the inflow of material (\S\ref{pulsar}). However, the situation changes in a significant way from that described until now. 

When the relative motions are fast (polar wind case, or high velocity isolated pulsar case), the standoff distance is farther out than the Bondi capture radius so that the external ram pressure is mostly unidirectional. The pulsar creates a long comet tail directed away from the maximum ram pressure point. At periastron, the standoff distance is within the Bondi capture radius. The external pressure from inflowing material is more spherically distributed, as in a SNR plerion, so that the shocked pulsar wind cannot break open the surrounding medium on a short timescale.

Therefore, a possible picture would be that the shocked pulsar wind is embedded in the equatorial disc at periastron (plerion-like) and that the geometry is open (comet-like) at apastron, where the Bondi radius is smaller than $R_s$ for both an equatorial or a polar wind. High energy radiation is not affected much as it is emitted close to the pulsar in both cases. The change should be mostly in the low frequency emission as particles are constantly injected in a bubble slowly expanding in the equatorial outflow. The outburst would be associated with emission from this bubble of electrons, following an earlier suggestion by \citet{leahy2004}. Detailed modelling is not attempted here but note that adiabatic expansion of bubbles has been shown to reproduce the radio outbursts \citep{taylor84,paredes91,leahy2004} and, naturally, has also been proposed for the radio outbursts of \psrb\ \citep{connors}.

\subsubsection{Radio map}
\begin{figure}
\resizebox{4.3cm}{!}{\includegraphics{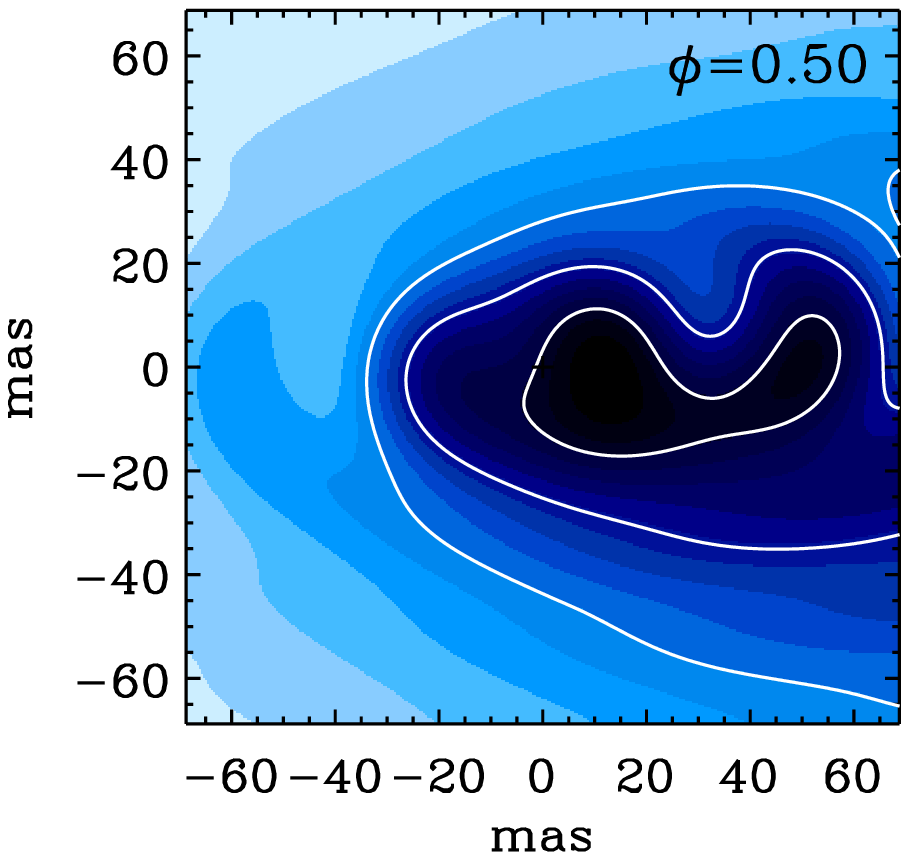}}
\resizebox{4.3cm}{!}{\includegraphics{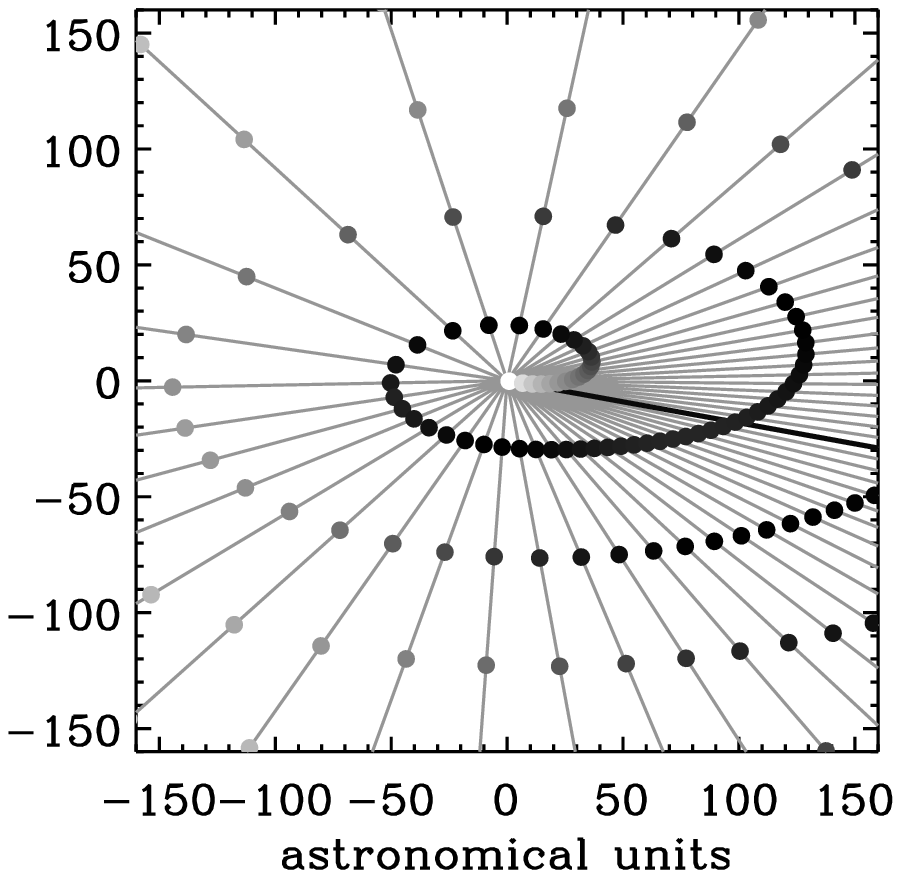}}
\caption{Left: Map of the 5~GHz radio emission from the pulsar nebula in \lsi\ at $\phi=0.5$ ($i=120$\degr). $\phi$ {is zero at periastron passage (for which} $\phi_{\rm radio}=$0.23, \citealt{gregory02}). Right: Corresponding outflow path (as in Fig.~\ref{spiral} {but on a linear greyscale}) before smoothing ({the projected outflow path at apastron is indicated by a dark line}). The stellar wind is assumed to be polar (top panel of Fig.~\ref{lsimodsed}). The smoothing Gaussian kernel has a FWHM of 8 mas (2.8 $10^{14}$ cm at 2.3~kpc). {Contours are at 50\%, 25\% and 10\% of maximum emission. There is a preferred direction to the right, towards apastron}. \label{lsiradio}}
\end{figure}

The radio emission map expected from the simple case of a polar wind is shown in Fig.~\ref{lsiradio} for $\phi=0.5$ and $i=120\degr$. (Here, $\phi=0$ corresponds to periastron passage, to which $\approx0.2$ should be added if using the radio ephemeris, \citealt{gregory02}.) The orientation of the orbit is illustrated in Fig.~2 of \citet{casares2}. The image does not take into account the radio outbursts but should nevertheless be indicative of the overall morphology to expect.

As in \ls, the nebula has a `comma' shape. The physical scale is larger than in \ls\ because of the longer orbital period. The major axis is almost in the plane of the sky, defining a very clear preferred direction for emission. This is consistent with observations of one-sided radio emission on scales of 10-100 mas \citep{massi01,Massi2004}. The mean position angle will increasingly vary as one goes to higher spatial resolutions and the impact of the orbital motion becomes clearer. This {could explain} the almost perpendicular directions seen at 3 mas and 10 mas resolution, albeit at different epochs \citep{massi93,massi01}. {\citet{Massi2004} also find changes in structure about 50 mas away from the core emission on timescale of days during the radio outburst. This could be due to a combination of rapidly changing emission angles at periastron passage and delayed peak radio emission; a proper model for the radio outbursts would be required to verify this possibility.}

\subsection{\psrb\label{psrb}}

For \psrb\, the measured $\dot{E}$ is 8$\cdot 10^{35}$~erg~s$^{-1}$ \citep{manchester95}. The other pulsar parameters were taken to be $\sigma=0.01$ and $\gamma_p=10^5$ to compare with the other two systems. Emission by containment of a pulsar wind in \psrb\ has previously been investigated by several authors \citep{tavani,melatos,ta97,kirk,murata}. The purpose here is to put the system in the same context as \ls\ and \lsi, and explore possible large scale radio emission.

 Both an equatorial and a polar stellar wind were considered, with the same parameters as those given for \lsi\ in \S\ref{wind}. The periastron distance in \psrb\ is comparable to the apastron distance in \lsi. Any equatorial outflow is then at or close to its terminal velocity and there is little difference in $R_s$ with a polar wind (only a factor 3, see Table~\ref{freq}). The equatorial wind is truncated at $\sim 15-20~R_\star$ and is not considered at apastron. Only a polar wind can contain the pulsar wind at apastron.

\subsubsection{The SED}
\begin{figure}
{\includegraphics[width=\columnwidth]{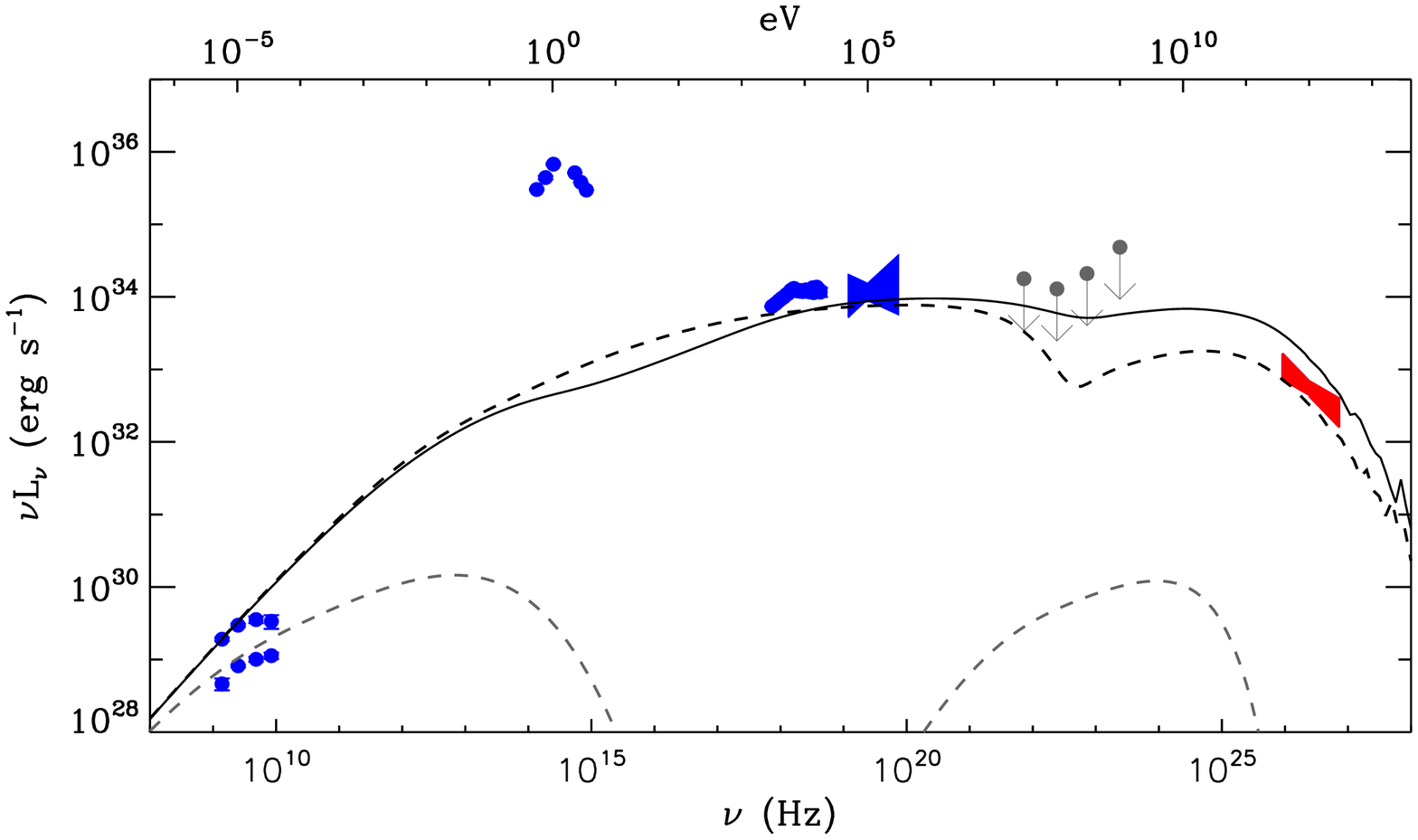}}
\caption{Spectral energy distribution of \psrb\ at periastron (solid line) and apastron (dashed line).   The pulsar is assumed to have $\dot{E}=8\cdot 10^{35}$~erg~s$^{-1}$, $\gamma=10^5$ and $\sigma=0.01$. The stellar wind is polar (same parameters as \lsi) with a standoff point at $R_s=3\cdot 10^{12}$~cm (periastron) and $4\cdot10^{13}$~cm (apastron). With an equatorial, slow wind, the standoff point at periastron is at $R_s=10^{12}$~cm, not different enough from the polar wind case to significantly change the expected SED from that shown above. The grey dashed line show the summed late time emission ($\Delta t> P_{\rm orb}$) at apastron, which occurs on scales $\ga 0.01$~pc. \label{psrbsed}}
\end{figure}

The spectral energy distribution at periastron and apastron is shown in Fig.~\ref{psrbsed} for the case of a polar wind. The SED at periastron for an equatorial wind (not shown) is very close to the one shown here, the difference in $R_s$ being small. Unsurprisingly, the SED is essentially the same as that for \lsi\ at apastron, the orbital separation being the same and other parameters varying little. GeV emission from \psrb\ around periastron should be detected by {\em GLAST} in the future, if only because of this similitude. The TeV slope appears in good agreement with the \hess\ measurement. The variations in integrated flux seen by \hess\ around periastron are beyond this model's present abilities (which predicts none) and will probably require taking into account anisotropic IC scattering \citep[see][]{kirk}, as well as changes in the circumstellar density to properly reproduce. There seems to be an inflection in the X-ray spectrum between 1-10~keV that might correspond to the break observed with {\em RXTE}. Very good agreement can indeed be obtained when using $\gamma_p=10^6$.

The SED at apastron is only slightly different. Despite the considerably larger orbital separation, the TeV flux is only lowered by factors of a few. The photon density is indeed much lower, but the scale of the flow (set by $R_s$) is much larger than at periastron. Particles have a longer time available to interact and this compensates for the weaker densities. Overall, the radiation efficiency is the same as at periastron as the same amount of energy is being dissipated regardless of orbital phase. The model cannot account in its present simplified form for the observed order-of-magnitude changes in X-ray luminosity. This was noted by \citet{murata} who proposed that lowering $\sigma$ with pulsar distance might help.

Emission at apastron occurs on large, resolvable scales. The grey dashed lines in Fig.~\ref{psrbsed} show the amount of the total flux emitted after an elapsed time $\Delta t > P_{\rm orb}$ after injection. This flux is radiated several {hundreds of} AUs away and may lead to a faint arcsecond scale bowshock structure up to IR frequencies. 

\begin{figure}
\resizebox{4.3cm}{!}{\includegraphics{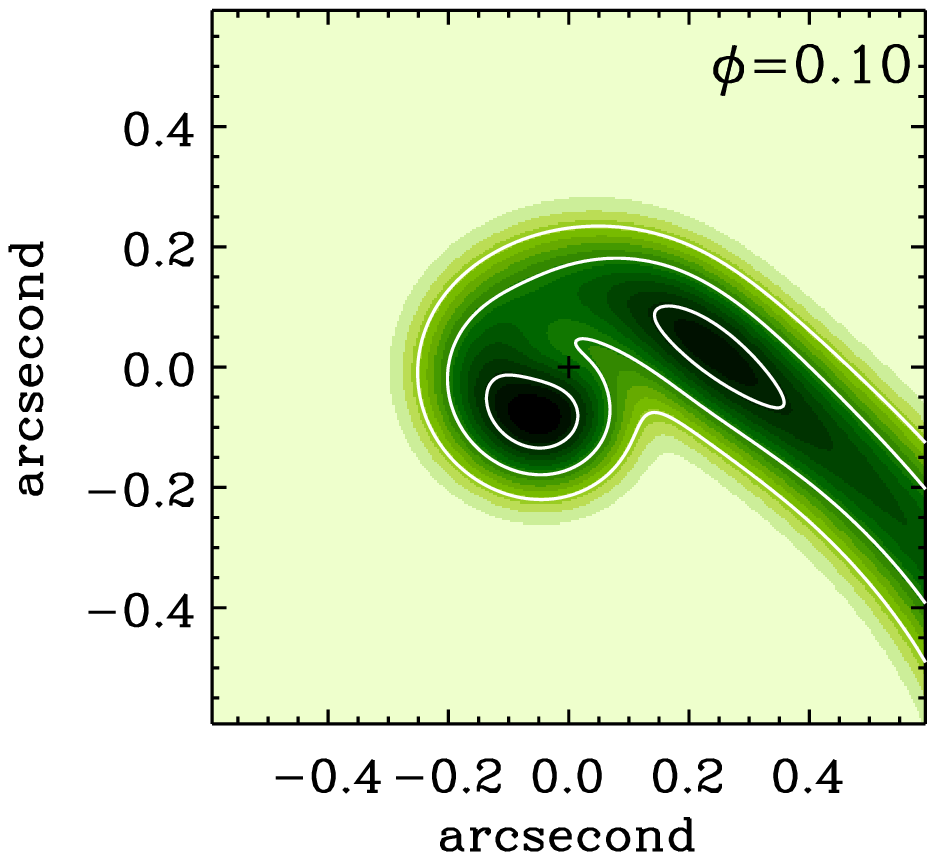}}
\resizebox{4.3cm}{!}{\includegraphics{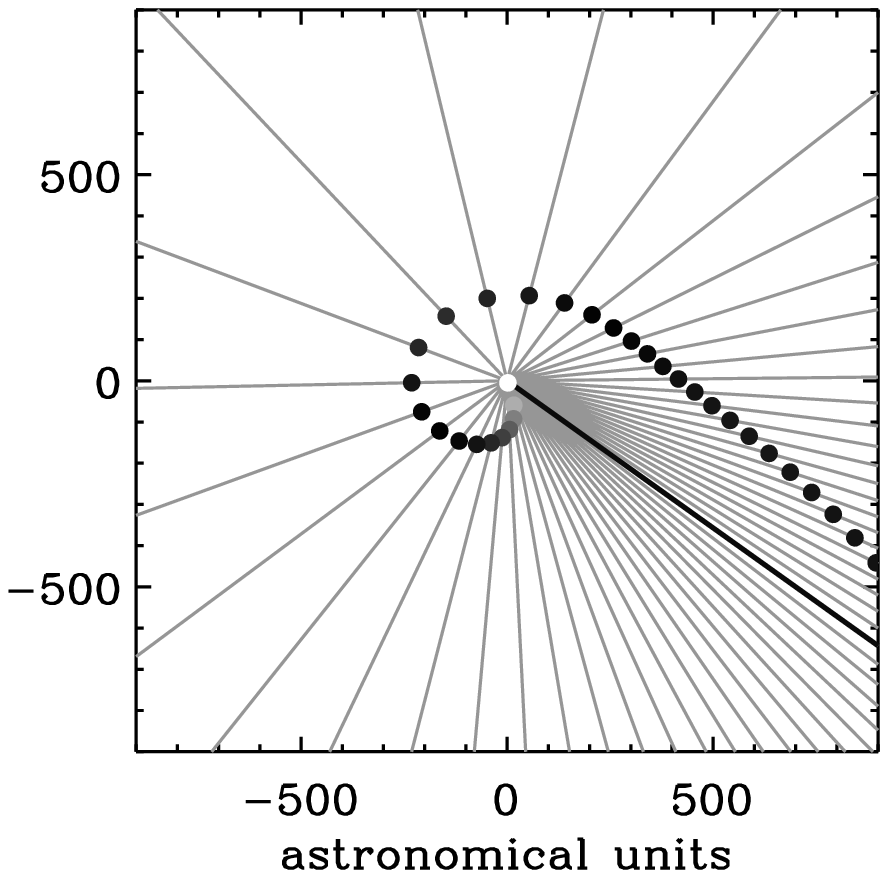}}
\caption{Left: Map of 5~GHz radio emission from \psrb\ close to periastron passage ($\phi=0.1$, $i=30$\degr), assuming only a polar-type stellar wind.  Right: Corresponding outflow path on the same scale before smoothing ({the projected outflow path at apastron is indicated by a dark line}). The smoothing gaussian kernel has a FWHM of 0.05\arcsec\ (1.1 $10^{15}$ cm at 1.5~kpc). The contours are at 50\%, 25\% and 10\% of the maximum intensity. The tail extends to a few arcseconds. The spatial scale is large given the size of the orbit (hence $R_s$); with an even larger eccentricity than \lsi, the preferred direction of the nebula towards apastron is very clear. \label{psrradio}}
\end{figure}

\subsubsection{Radio outbursts and map}
In \psrb, observations of the radio dispersion measure imply the disc is inclined with respect to the orbital plane \citep{melatos}. The radio outbursts seen before and after periastron are presumably associated with the two pulsar disc crossings \citep{ball}. Working by analogy, smaller separations and a coplanar equatorial outflow would explain the single outburst per orbit in \lsi. As explained above for \lsi, the outbursts are unlikely to be due solely to a change in $R_s$. The large increase in radio flux might be related to the slow expansion of a bubble of relativistic electrons formed in the dense flow. Shock-acceleration of electrons from the stellar wind itself may also play a role \citep{ball}.

It is therefore as difficult as with \lsi\ to accurately predict the shape the radio emission might take. Fig.~\ref{psrradio} shows expectations at $\phi=0.1$ assuming \psrb\ only interacts with a polar wind. The spatial scales should be easy to resolve since $R_s$ is large. Unfortunately there are no published radio maps to compare with. The rapid revolution of the pulsar around periastron produces an arc with two brighter `knots' where pre- and post-periastron electrons are cooling. These two knots gradually drift apart. This takes no account of the radio outbursts. If these are localised in the equatorial wind, the impact might be only an increase in the central (unresolved) component, leaving the much fainter diffuse components described here mostly unchanged. The bow shock points away from the star in almost the same direction for most of the highly eccentric orbit.

\section{Conclusion}

The pulsar wind scenario offers a tantalizing way to unite \ls, \lsi\ and \psrb\ into a new-class of young, rotation-powered $\gamma$-ray binaries. At present, nothing rules out that the compact object in \ls\ and \lsi\ is a pulsar. The radial velocities allow it and there are no {conclusive} signs of accretion. Directly unveiling a radio pulse will not be possible due to strong free-free absorption; although a difficult endeavour, an X-ray or $\gamma$-ray pulse would prove the scenario.

A similar relativistic pulsar wind with most of its $\dot{E}\approx 10^{36}$~erg~s$^{-1}$ energy ($\sigma\approx 0.01$) in electrons and positrons of Lorentz factor $\gamma_p\approx 10^5$ suffices to explain most properties in all three objects. The termination shock occurs at the pressure balance point. There, particles in the pulsar wind are scattered and accelerated into a non-thermal power-law distribution. The maximum energy is set by the size of the shock region and radiative losses. Downstream values in the shocked pulsar wind material (density, magnetic field) are set by the MHD jump conditions. Particles radiate synchrotron and inverse Compton whilst being advected away in a comet tail nebula. At high energies, synchrotron dominates over the Klein-Nishina cross-section, giving characteristic break frequencies between a rising and flat $\nu F_\nu$ spectrum in hard X-rays, and between a flat and a declining $\nu F_\nu$ spectrum in VHE $\gamma$-rays. {The VHE emission occurs close to the pulsar and an orbital modulation of the flux in the TeV range is expected in this scenario, due to absorption (pair production) on stellar photons. }

{The small-scale radio emission is also reasonably reproduced}. Radiation shifts to lower wavelengths as particles lose their energy to radiation and conditions in the flow change. Ultimately, the particles that gave rise to $\gamma$-ray radiation near the star emit radio far from the binary. The flow changes from a cometary shape to a spiral-shaped nebula with a step $\sim \sigma c P_{\rm orb}$. Matching with the observed size of radio nebula in \ls\ and \lsi\ implies $\sigma \approx 0.01$. {On large scales (several turns of the spiral), the radio emission is underestimated by the model. Changes in the flow conditions on such scales, notably velocity and magnetic field evolution, are undoubtedly more complex than the simple assumptions made here. Nevertheless,} the combination of the peculiar geometry, emission at various distances and inclination can conspire to create radio maps with seemingly bipolar ejections of material {as in X-ray binary jets}. Predictions are made for \ls\ that are amenable to observation with radio interferometry.

The {present} model does not provide an entirely satisfactory view of the GeV $\gamma$-rays detected by EGRET in \ls\ and \lsi. The fluxes are in both cases underproduced. The discrepancy might have an observational side, with fluxes overestimated,  comprising the combined emission from several sources in the wide instrumental PSF. Much more precise measurements will be made by {\em GLAST} in the future. On the theoretical side, inadequacies in the treatment of the emission region might be at work, notably anisotropies. The stellar radiation is anisotropic, the pulsar wind is anisotropic, the termination shock is not spherical etc. If synchrotron losses are limited, IC cooling of particles will enhance the GeV yield and lower the discrepancy. The mismatch between EGRET and HESS spectra of \ls\ is also difficult for {leptonic} jet models \citep{dermer2,paredes05}. 

The {present} model does not either satisfyingly explain the nature of the radio outbursts in \lsi\ and \psrb. However, the scenario provides {a unifying context}. As has been argued {for both sources}, the reason must be found in the Be nature of the {companion stars}. The progression of the pulsar in the slow, dense wind could lead to an enclosed geometry, rather than an open comet tail geometry. Radio emission might also be enhanced by the acceleration of particles from the stellar wind itself. There is as yet no coherent picture of the radio outbursts of \lsi\ in the context of the accretion-driven / relativistic jet scenario.

Despite these shortcomings, the explanatory scope of the model is large. Once the orbital parameters are set, the only variables that come into play are $\dot{E}$, $\sigma$, $\gamma_p$ and $R_s$.  {The SEDs come out naturally for reasonable inputs and the comet-tail nebula produces radio emission on scales of $\sigma P_{\rm orb} c$ and beyond. The pulsar scenario therefore seems like a viable alternative to the accretion/jet scenario for \ls\ and \lsi.}

\begin{acknowledgements}
I thank B. Giebels, J.-P. Lasota, M. Sikora, R. Taam and R. Fender for their comments on this work.
\end{acknowledgements}

\end{document}